%% file: main.tex
\documentclass[acmlarge]{acmart}
\pdfoutput=1%For ARXIV
\makeatletter
\def\@ACM@checkaffil{% Only warnings
    \if@ACM@instpresent\else
    \ClassWarningNoLine{\@classname}{No institution present for an affiliation}%
    \fi
    \if@ACM@citypresent\else
    \ClassWarningNoLine{\@classname}{No city present for an affiliation}%
    \fi
    \if@ACM@countrypresent\else
        \ClassWarningNoLine{\@classname}{No country present for an affiliation}%
    \fi
}
\makeatother

\AtBeginDocument{%
  \providecommand\BibTeX{{%
    \normalfont B\kern-0.5em{\scshape i\kern-0.25em b}\kern-0.8em\TeX}}}

\setcopyright{acmlicensed}
\acmJournal{IMWUT}
\acmYear{May2024}
\acmDOI{11.1111/1111111}

\acmISBN{978-1-4503-XXXX-X/18/06}

\usepackage{subfigure}
\usepackage{amsmath}
\usepackage{algorithm}
\usepackage{algorithmicx}
\usepackage{algpseudocode}
\usepackage{graphicx}
\usepackage{lipsum}
\usepackage{enumerate}
\usepackage{enumitem}
\usepackage{hyperref}
\usepackage{multirow, makecell}
\usepackage{nccmath}
\usepackage{ragged2e}
\usepackage{threeparttable} 
\usepackage{url}
\usepackage{xspace}
\usepackage{caption}
\usepackage{subcaption}
\usepackage{bbding}
\usepackage{graphicx}
\usepackage{subcaption} % 添加子图标题
\usepackage{lipsum} % 添加示例文本
\usepackage{CJKutf8}

\newcommand{\ie}{\emph{i.e.},\xspace}

\newcommand{\eg}{\emph{e.g.},\xspace}
\newcommand{\etc}{\emph{etc.}\xspace}

\newcommand\figref[1]{Fig.~\ref{#1}}

\newcommand\tabref[1]{Tab.~\ref{#1}}

\newcommand\secref[1]{Sec.~\ref{#1}}

\newcommand\equref[1]{Equ.(\ref{#1})}

\newcommand{\sysname}{{\sf UbiHR}\xspace}

\newcommand\re[1]{\textcolor{black}{#1}}
\ifodd 0
\newcommand{\TODO}[1]{\textbf{\color{red}{TODO: #1} }}

\newcommand\lsc[1]{\color{black}{#1}}
\newcommand\bhy[1]{\textcolor{cyan}{#1}}

\else
\newcommand{\TODO}[1]{#1}

\newcommand\lsc[1]{#1}
\newcommand\bhy[1]{#1}

\fi

\begin{document}
\begin{CJK}{UTF8}{gbsn}

\title{UbiHR: Resource-efficient Long-range Heart Rate Sensing on Ubiquitous Devices}

\author{Haoyu Bian}
\orcid{0009-0004-5432-8965}
\affiliation{%
  \institution{Northwestern Polytechnical University}
  \department{School of Computer Science}
  \city{Xi'an}
  \country{China}
}

\author{Bin Guo}
\orcid{0000-0001-6097-2467}
\affiliation{%
  \institution{Northwestern Polytechnical University}
  \department{School of Computer Science}
  \city{Xi'an}
  \country{China}
  \thanks{Corresponding author: 
    Bin Guo(guob@nwpu.edu.cn), Sicong Liu(scliu@nwpu.edu.cn)}
}

\author{Sicong Liu}
\orcid{0000-0003-4402-1260}
\affiliation{%
  \institution{Northwestern Polytechnical University}
  \department{School of Computer Science}
  \city{Xi'an}
  \country{China}
}

\author{Yasan Ding}
\orcid{0000-0001-9051-5865}
\affiliation{
  \institution{Northwestern Polytechnical University}
  \department{School of Computer Science}
  \city{Xi'an}
  \country{China}
}

\author{Shanshan Gao}
\orcid{}
\affiliation{
    \institution{Hospital of Xi'an Jiaotong University}
    \department{Department of Cardiovascular Medicine}
    \city{Xi'an}
    \country{China}
}

\author{Zhiwen Yu}
\orcid{0000-0002-9905-3238}
\affiliation{
    \institution{Northwestern Polytechnical University, Harbin Engineering University}
    \department{School of Computer Science}
    \city{Harbin}
    \country{China}
}

\begin{abstract}
Ubiquitous on-device heart rate sensing is vital for high-stress individuals and chronic patients. 
Non-contact sensing, compared to contact-based tools, allows for natural user monitoring, potentially enabling more accurate and holistic data collection. 
However, in open and uncontrolled mobile environments, user movement and lighting introduce \re{noises}.
Existing methods, such as curve-based or short-range deep learning recognition based on adjacent frames, strike the optimal balance between real-time performance and accuracy, especially under limited device resources.
In this paper, we present \sysname, a ubiquitous device-based heart rate sensing system. 
% \sysname incorporates an energy-efficient adaptive duty-cycling video sampling module, a dynamic noise-aware facial i preprocessing module, and a fast and multi-branch long-range spatio-temporal recognizing model.
Key to \sysname is a real-time long-range spatio-temporal model enabling noise-independent heart rate recognition and display on commodity mobile devices, along with a set of mechanisms for prompt and energy-efficient sampling and preprocessing. 
Diverse experiments and user studies involving four devices, four tasks, and 80 participants demonstrate \sysname's superior performance, enhancing accuracy by up to 74.2\% and reducing latency by 51.2\%.
\end{abstract}




\received{1 May 2024}

\maketitle

\input{body/1-Introduction}
\input{body/2-design-research}

\input{body/3-system-design}

\input{body/4-implementation}
\input{body/5-evaluation}
\input{body/6-related-works}

\input{body/conclusion}

\bibliographystyle{ACM-Reference-Format}
\bibliography{reference}

\input{body/appendix}

\end{CJK}
\end{document}

%% file: body/1-Introduction.tex
\section{INTRODUCTION}
Physiological health sensing anytime and anywhere is important for users, encompassing metrics like heart rate~\cite{palatini1997heart,fox2007resting}, blood oxygen~\cite{geisler2006blood}, blood pressure~\cite{frattola1993prognostic}\re{, \etc}
Existing methods include hospital-grade and consumer-grade solutions, both contact-based and non-contact. 
However, discomfort is a common issue with contact-based options like specialized vests~\cite{kiczek2022lets} and fingertip monitors~\cite{hashem2010design}, while wearable devices such as watches~\cite{kawasaki2021estimating}, wristbands worn on the wrist~\cite{wang2022emotracer,alchieri2022impact,cao2022guard}, and glasses~\cite{chwalek2021captivates} impose limitations by compressing the skin and affecting daily activities.
There is a need for a \textit{ubiquitous}, \textit{non-contact} heart rate sensing tool utilizing mobile and embedded cameras. 
This allows for user monitoring in \textit{natural settings}, leading to potentially more accurate and holistic data
collection.
For instance, camera-based heart rate sensing can aid law enforcement in detecting truthfulness during confessions without awareness or psychological resistance, and enabling convenient health monitoring for elderly individuals living alone.

Meanwhile, the pervasive deployment of camera-embedded devices \eg smartphones, wearables, tablets, and robots has stimulated a wide spectrum of novel video-based physiological applications. 
% \TODO{cite application papers here}
Examples include Mobilephys on smartphones~\cite{liu2022mobilephys} and TS-CAN on embedded platforms~\cite{liu2020multi}.
Despite extensive research on camera-based heart rate sensing \cite{xiao2024remote,verkruysse2008remote,chen2018deepphys,yu2019remote,niu2019rhythmnet,yu2022physformer}, they are unfit for deployment to ubiquitous mobile and embedded devices because they fail to meet the following requirements.

\begin{itemize}
    \item \textit{Low-cost} \textit{end-to-end} heart rate sensing on resource-constrained ubiquitous devices is non-trivial for practical applications. 
    \re{While video-based remote photoplethysmography (rPPG) measurements are commonly used for heart rate sensing, privacy concerns dictate \textit{on-device preprocessing and processing}~\cite{li2024echopfl} instead of cloud streaming. }
    However, enabling this capability on low-end mobile and embedded devices faces challenges due to high computing overhead. 
    Previous studies rely on hand-crafted methods~\cite{chen2018deepphys} or complex pre-processing techniques, such as retaining only the facial region of interest (RoI) through face detection~\cite{gudi2019efficient}, dividing the image into small blocks/channels~\cite{tran2015robust,lu2021dual}, or constructing a 3D model~\cite{li2023learning}, assuming given pre-processed data. 
    As demonstrated in \secref{sec:evaluation}, these methods struggle to realize high accuracy with low on-device latency.
    
    \item
    \textit{Real-time} spatio-temporal heart rate sensing from noisy video streams poses a significant challenge. 
    Unlike traditional analysis of signals acquired by contact-based devices, which focuses on prominent changes, facial rPPG measurement requires capturing subtle facial movements and color variations, complicating global spatial-temporal recognition. 
    Also, adapting existing curve-based models, such as color space projection~\cite{botina2022rtrppg, liu2023efficientphys} and skin reflection~\cite{wang2016algorithmic}, to dynamic open environments \re{(\eg various light conditions~\cite{liu2023adaenlight})} proves challenging, as none of them consider rPPG signal periodic properties in long-range sensing~\cite{yu2022physformer}. Factors such as user head motions, lighting fluctuations, and equipment noise further complicate the recognition process. Addressing these challenges necessitates intricate long-range and noise-robust spatio-temporal DL models, which are unsuitable for resource-constrained ubiquitous devices.
    
\end{itemize}

In this paper, we present \sysname, a real-time and continuous \textit{\re{on-device}} heart rate sensing system for ubiquitous mobile and embedded devices with \re{\textit{long-range}} video streams.
To ensure user-friendly design and reduce bias, we conduct pre-design surveys with 80 participants, including 30 students and 50 patients, to understand user demands.
Achieving real-time and high-accuracy performance demand with non-stationary noisy videos on low-end resource-limited devices poses significant challenges. 
Our \textit{key insight} is that detecting on long-range is more conducive to detecting the inherent periodic characteristics and trends of heart rate, enhancing the ability to resist environmental noise.
% We propose long-range spatio-temporal modeling to enhance accuracy and noise robustness.
Specifically, \sysname incorporates three modules. 
\textit{First}, an \textit{adaptive duty-cycling facial video sampling} module dynamically switches between sleeping, short-term listening, and long-term listening schemes to balance energy consumption and promptness of facial key point detection. 
\textit{Second}, a \textit{dynamic noise-aware facial image pre-processing} module extracts essential facial key points through differentiation and normalization, eliminating noise from user head motion and varying environmental lighting conditions while reducing input volume for subsequent real-time processing. 
\textit{Third}, a \textit{fast long-range spatio-temporal heart rate recognizing} module integrates a zero-parameter time shift strategy into 2D convolution for lightweight spatio-temporal modeling. 
It extracts both long-range and short-range spatio-temporal features from the heart rate signal using parallel multi-branch structures and corrects accuracy using soft-attention masks. 

We implement \sysname as a python package that can be easily integrated into other applications on four typical mobile and embedded devices, \ie IQOO9 smart phone, Raspberry, Jetson orin, and Jetson AGX.  
We evaluate the performance of \sysname on four open datasets and real-world scenarios with diverse user motion or light conditions.
Results show a reduction of up to 74.2\% in recognizing error (\ie MAE, MAPE, RMSE), up to 51.2\% in inference latency, 49\% in memory cost, and 19\% in energy cost. 
Especially, \sysname achieves up to 42.8\% accuracy increase in long-range video streams (\secref{sec:4.5}).
Our main contributions are summarized as follows:

\begin{itemize}
    \item To the best of our knowledge, \sysname is the first \textit{real-time} \textit{on-device} \re{contactless} heart rate sensing tool on ubiquitous devices with \textit{long-range} video streams. 
    It delivers real-time, high-accuracy video sampling, preprocessing, and processing on commodity platforms while maintaining robustness against various noise.
    
    \item We realize \sysname as an end-to-end mobile system. 
    The key technical novelties include adaptive duty-cycling facial video sampling, dynamic noise-aware facial image preprocessing, and fast long-range spatio-temporal heart rate recognizing.
    \item We evaluate \sysname on 4 public benchmarks and real-world ubiquitous scenarios on four mobile platforms.
    Experiments show that \sysname achieves near real-time (5.01 $\sim$ 30.9 ms) with over 20.9\% higher accuracy than state-of-the-art methods.
\end{itemize}

In the rest of this paper, we conduct a pre-design survey in \secref{sec:2},
detail the system design in \secref{sec:design}.
We next present the system implementation and evaluate the system in \secref{sec:evaluation}.
Finally, we review the related work in \secref{sec:related} and conclude in \secref{sec:conclusion}.

%% file: body/2-design-research.tex
\section{\lsc{User study}}
\label{sec:2}
\label{sec:study}

\begin{figure*}
\centering
\includegraphics[width=0.5\textwidth]{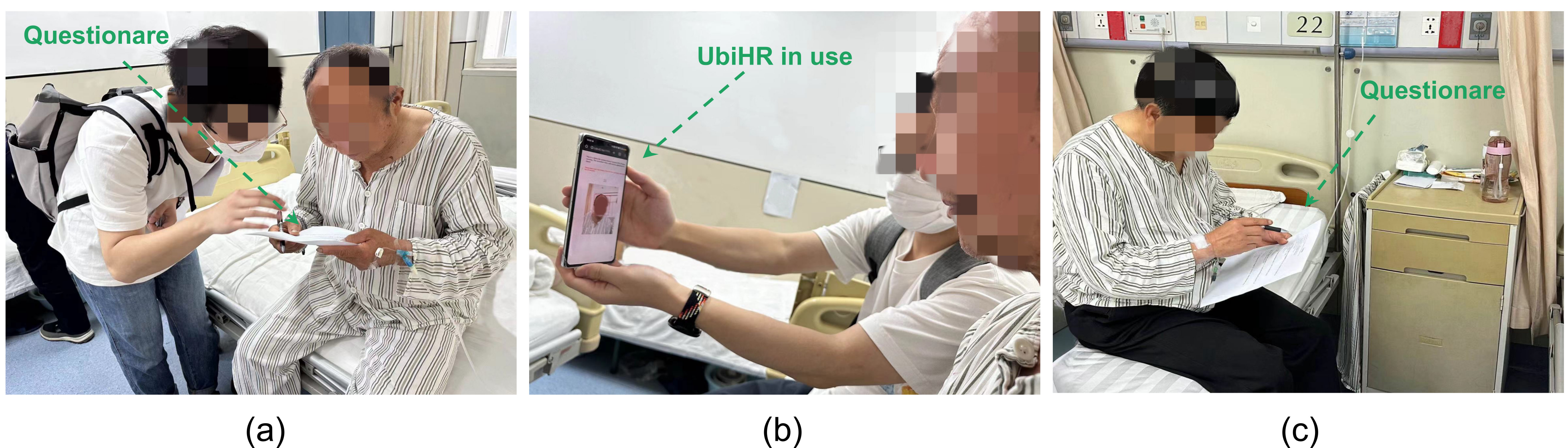}
\caption{Questionnaire survey on a hospital cardiology department and user experience of UbiHR.}
\label{fig_survey}
\end{figure*}

\lsc{To understand the needs and usage preferences of ubiquitous users regarding a camera-based heart rate sensing tool, we designed a questionnaire and distributed it to 80 users from our school and a hospital cardiology department.
Participants' ages ranged from 20 to 76 years old, as shown in Fig. \ref{fig_survey} 
Permission was obtained from both the hospital and participants, and the questionnaire included multiple-choice and open-ended questions.
}

\subsection{Survey Findings}
\begin{figure*}
\centering
\includegraphics[width=0.9\textwidth]{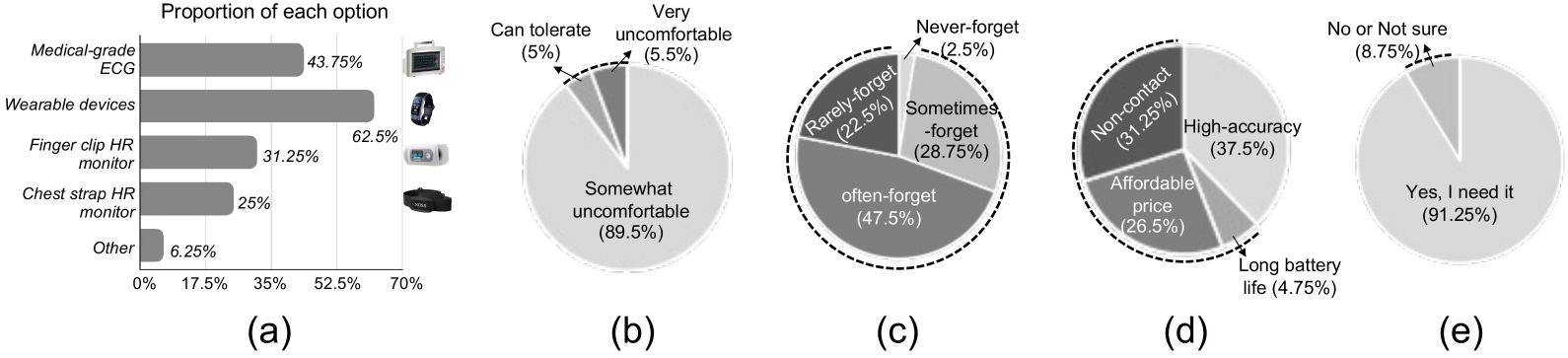}
\caption{Survey results: (a) What method(s) do you currently use to measure HR? (b) Do you feel uncomfortable with the HR measurement method(s) that require skin contact (such as wearing a chest strap or finger clip?) (c) Do you often forget to wear or use HR measurement devices that require skin contact? (d) What features would you like the tool to have? (e) Do you think there is a need for a contactless HR sensing tool in addition to your current HR measurement method(s)?}
\label{fig_survey_findings}
\end{figure*}

We briefly summarize the results (as shown in Fig. \ref{fig_survey_findings}) of our survey as follows.

\subsubsection{User Demands on Fast and Long-term Heart Rate Sensing}
\lsc{
Through surveys, various groups of individuals require frequent and long-term heart rate sensing:
\textit{i)} Patients with cardiovascular, hypertension, and diabetes conditions need it to assess treatment effectiveness, evaluate blood pressure control, assess cardiovascular risk, and prevent complications.
For instance, individuals with arrhythmia need continuous monitoring to assess treatment effectiveness, evaluate blood pressure control, assess cardiovascular risk, and prevent complications. 
Long-term heart rate sensing aids in the timely detection and treatment of potential heart issues, \re{mitigating the risk of cardiac events, \eg \textbf{too fast} (ventricular tachycardia), \textbf{too slow} (high degree atrioventricular block), or \textbf{irregular} (paroxysmal atrial fibrillation).}
\textit{ii)} Elderly users require it to assess cardiovascular health and promptly detect and manage potential issues. \re{As people age, the risk of cardiovascular diseases in the elderly rises, making timely cardiovascular health assessments crucial. 
It helps doctors to timely perform further examinations and diagnoses, facilitating early detection and treatment of potential cardiovascular issues.
}
\textit{iii)} Athletes and fitness enthusiasts need it to continuously evaluate exercise intensity and optimize training plans to prevent overtraining. \re{Unlike traditional methods, contactless heart rate monitoring eliminates the need for athletes to wear contact-based devices, ensuring their movements and performance during training remain unaffected.
}
\textit{iv)} Individuals managing stress and mental health benefit from continuous heart rate monitoring, as heart rate closely correlates with stress levels and emotional states. 
This is particularly important for young users experiencing high work stress who require continuous assessment of stress levels. 
\re{
This can boost users' self-awareness and motivate proactive relaxation and adjustment.
}
}

\subsubsection{Limitations of existing heart rate sensing tools.} 
\lsc{The majority of participants currently depend on wearable devices and medical-grade electrocardiogram (ECG) equipment at hospitals, constituting 62.5\% (50/80) and 43.75\% (35/80), respectively. 
However, participants voiced concerns: 
95\% of participants expressed discomfort with ECGs attached to the chest or clipped to fingers. 
\re{
This suggests that current medical-grade, contact-based ECG equipment may not offer a comfortable user experience, potentially affecting user acceptance and long-term usage.}
Furthermore, 76.25\% of participants admitted to frequently or occasionally forgetting to wear them, potentially compromising the long-term waiting time for health sensing. 
\re{
This compromises long-term data acquisition for health monitoring and detecting health trends or early warning signs.}
}

\subsubsection{Willingness to use \textit{contactless} and \textit{ubiquitous} heart rate sensing tools} 
\lsc{91.25\% of participants expressed the need for an additional heart rate sensing tool in addition to their current use of devices attached to the chest or clipped to fingers. 
Moreover, the desired features for this tool include high measurement accuracy (37.5\%) and a contactless sensing mode (31.25\%) that does not require attachment to the chest or finger.
}

\subsection{Design Goals}
\lsc{To further understand the requirements of the contactless heart rate sensing tool, we asked each participant to rate the importance of different performance aspects of the tool and summarized the results as our design goals.}
% \TODO{}
\begin{itemize}
    \item \textit{On-device}. 
    \lsc{60\% of participants are unwilling to stream out their sensitive raw videos to the cloud for analysis. 
    Thus, video stream collection and inference should be conducted locally on the device.
    \re{Additionally, 70\% of participants prefer interacting with local devices over the cloud to protect sensitive data privacy and reduce network dependency.}
    }
    \item \textit{Accurate}. 
    \lsc{80\% of participants reported low tolerance for errors, and 63\% of participants can only tolerate a maximum error of 5 times in terms of bmp (heart rate beats per minute).}
    % lsc: 误差不超过多少（阈值）
    \item \textit{Responsive}. 
    \lsc{70\% of participants are comfortable receiving results within 30ms of detection initiation.
    % lsc: xx 完整准确的判断
    Heart rate sensing and recognition should offer \textit{real-time performance} and prompt user notifications.
    }
    \item \textit{Energy-efficient}. 
    \lsc{Over 85\% of participants expressed concerns about the energy consumption of the tool on battery-powered ubiquitous devices, such as smartphones. 
    It should enable \textit{continuous sensing} with minimal energy cost.
    }
\end{itemize}

\subsection{Why Camera-based Physiological Sensing}
\label{subsec:principle}
\begin{figure*}
\centering
\includegraphics[width=0.82\textwidth]{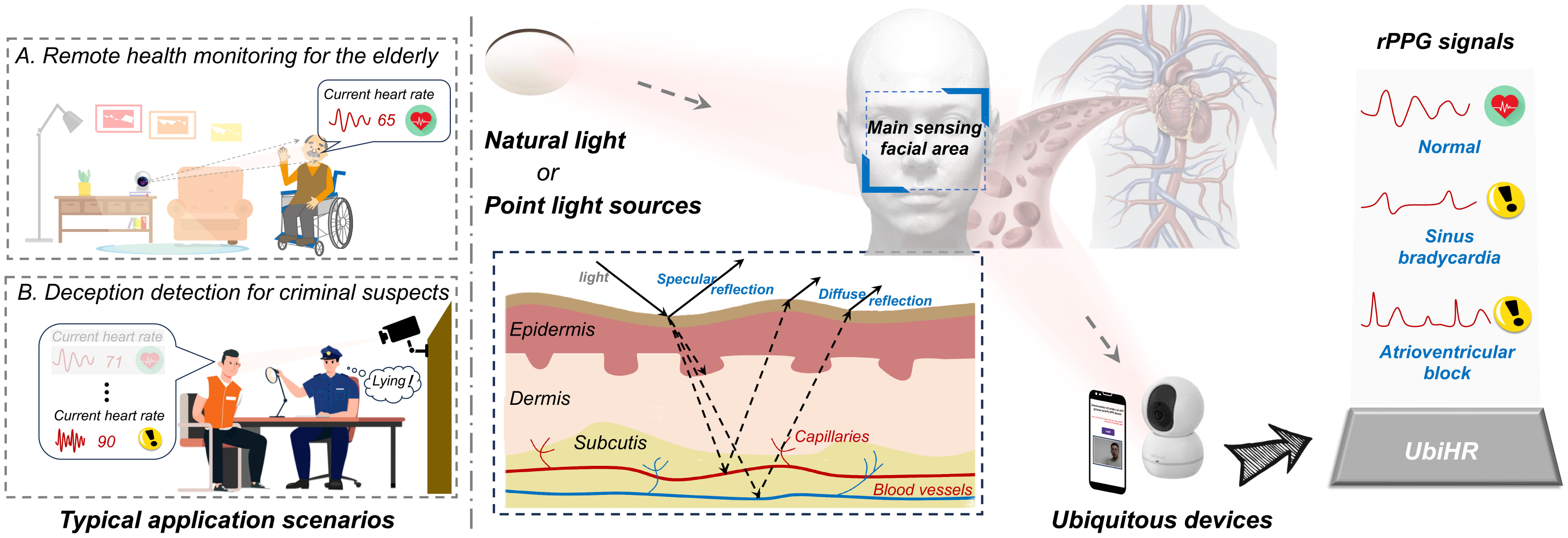}
\caption{\lsc{Illustration of the principle of mapping cardio-pulmonary and facial signal for contactless heart rate sensing.}}
\vspace{-3mm}
\label{fig_sensing_mechanism}
\end{figure*}

\lsc{Cameras can detect physiological signals by capturing facial videos, including heart rate, respiration rate, and blood oxygen saturation~\cite{scully2011physiological}.
This capability relies on the principle of contactless cardiopulmonary measurement, which involves detecting subtle changes in the reflected light from the body resulting from physiological processes.
As shown in Figure~\ref{fig_sensing_mechanism}, biologically, when blood volume increases (such as during each heartbeat), there is a corresponding increase in light absorption by the skin. 
This increase in blood volume leads to greater absorption of light. 
Consequently, the amount of visible light reflected by the skin changes, forming the basis of photoplethysmography (PPG) signals.
Camera-based imaging methods leverage these changes in light absorption to measure variations in blood volume on the skin surface. 
By analyzing the color and motion changes in the captured video, it becomes possible to extract pulse signals and calculate heart rate frequency.}

\re{The ubiquitous camera serves as a \textit{contactless} tool to monitor physiological parameters \textit{at any location and any time}.
This is beneficial for three reasons}:
\textit{i)} Compared to hospital, \re{contact-based} sensing methods such as electrocardiography (ECG) and pulse oximetry, ubiquitous camera-based sensing eliminates direct body contact, reducing discomfort and infection risks.
\textit{ii)} ubiquitous camera-based sensing seamlessly integrates into environments, enabling \textit{continuous sensing}. 
\re{
Users can be monitored in their \textit{\textbf{natural settings}}, leading to potentially more \textbf{accurate and holistic data collection}, compared to limited snapshot measurements in clinical settings. 
\textit{iii)} Commercial smart devices such as smartphones, laptops, and robots are widely equipped with cameras, making this contactless monitoring method highly accessible and scalable across a wide range of user populations and application scenarios.
We identify two application scenarios. Deception detection, such as in criminal suspect lie detection~\footnote{Heart rate-based deception detection deployment requires local authorization.}, where contactless camera-based heart rate sensors test truthfulness \textit{without the subject's awareness or psychological resistance}~\cite{duran2018resting, chang2023noncontact}. 
Another application is health monitoring, providing elderly people living independently with a convenient method to monitor their physical data and health status daily.
}

%% file: body/3-system-design.tex
\section{SYSTEM DESIGN}
\label{sec:design}
\lsc{This section begins with an overview of \sysname before elaborating on its design details. }

\subsection{Overview}
\begin{figure}[t]
    \centering
    \includegraphics[width=0.9\textwidth]{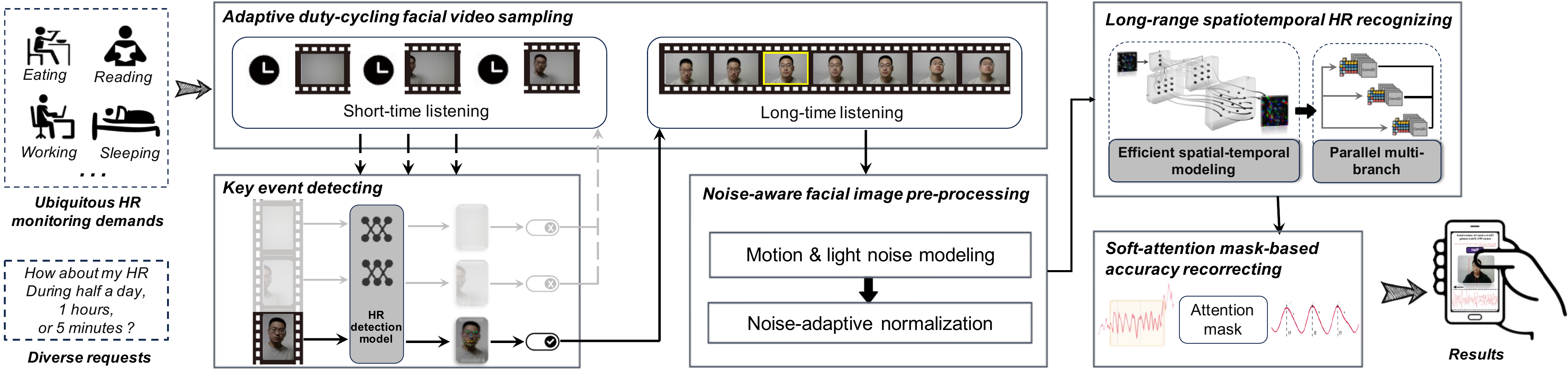}
    \caption{\lsc{Workflow of \sysname system.}}
    \vspace{-3mm}
    \label{fig:workflow}
\end{figure}

\lsc{Figure \ref{fig:workflow} shows the workflow of \sysname, which consists of four functional modules, \ie an \textit{adaptive duty-cycling facial video sampling} module, a \textit{noise-aware facial image preprocessing} module, and a \textit{fast long-range spatiotemporal heart rate recognizing} module.
The \textbf{adaptive duty-cycling facial video sampling} module switches between sleeping, short-term listening, and long-term listening schemes.  
It adopts an event-driven adaptive duty-cycling mechanism to balance the energy consumption and the promptness of facial keypoint detection.
The \textbf{dynamic noise-aware facial image preprocessing} module extracts essential facial key points through differentiation and normalization, eliminating noise from user head motion and varying environmental lighting conditions, and reducing input volume for subsequent processing with real-time performance.
}
The \textbf{fast long-range spatio-temporal heart rate recognizing} module incorporates the zero-parameter time shift strategy to 2D convolution for lightweight spatio-temporal modeling, extracting both long-range and short-range spatio-temporal features from the heart rate signal via parallel multi-branch structures, and corrects accuracy using soft-attention masks.

\subsection{Adaptive Duty-cycling Facial Video Sampling}
% 不用用户触发，而是自动自适应采样

\re{
This module balances energy cost with efficiency of facial keypoint detection. 
Continuously operating the camera on battery-powered devices like robots and smartphones consumes significant energy, which is impractical~\cite{liu2021adaspring,liu2020adadeep}. Given that major heart rate events like sinus bradycardia and high-degree atrioventricular block are rare and brief, \sysname uses an adaptive duty-cycling mechanism to minimize camera runtime and energy use. 
It operates in three camera states: \textit{short-term sampling} to detect complete facial images and significant heart rate events, \textit{sleeping} to save energy, and \textit{long-term sampling} for continuous video collection when events are detected.
}
% This module balances the energy consumption and the promptness of facial keypoint detection.
% Continuous operation of the camera to collect facial video streams consumes a significant amount of energy, impractical for battery-powered ubiquitous devices like robots and smartphones~\cite{liu2021adaspring,liu2020adadeep}. }
% Since significant heart rate-related events, such as sinus bradycardia and high-degree atrioventricular block, occur infrequently and for short durations \re{in daily life}, \sysname implements an adaptive duty-cycling mechanism to reduce camera runtime and, consequently, energy consumption. 
% Specifically, \sysname employs three camera states: short-term sampling, long-term sampling, and sleeping. In \textit{short-term sampling} mode, video segments are sampled to detect sufficiently complete facial images and significant heart rate-related events.
% If no significant heart rate-related event is found, the system switches to \textit{sleeping} mode to conserve energy. 
% Otherwise, it enters long-term capture mode, continuously collecting video streams for further processing. 

\begin{figure}[t]
\centering %图片全局居中
\includegraphics[width=0.6\textwidth]{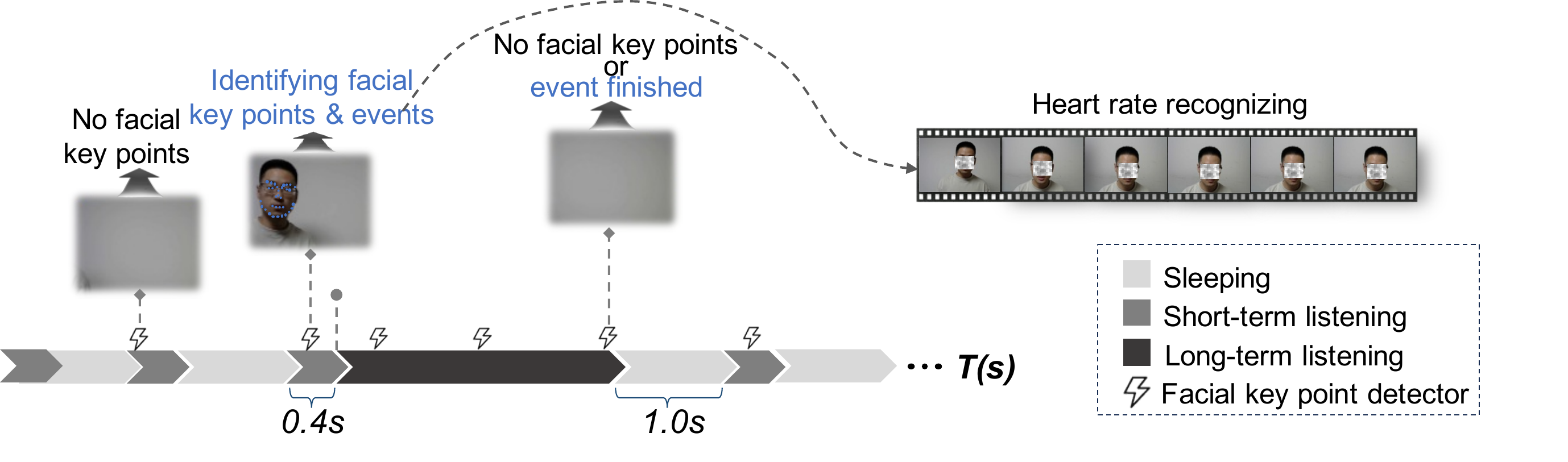} 
\vspace{-3mm}
\caption{\lsc{Illustration of the adaptive duty cycling facial video sensing mechanism (the mosaic is to protect privacy)}.}
\vspace{-3mm}
\label{fig:duty-cycling}
\end{figure}

\re{
Figure \ref{fig:duty-cycling} illustrates an \textit{event-driven}, \textit{adaptive} duty-cycling mechanism. 
Initially, the camera starts in \textit{short-term sampling} mode, adjusting for user motion in open environments. An event detector identifies valid video frames and heart rate-related events, shifting to long-term sampling if needed. 
Studies like \cite{irani2014improved} and \cite{moreno2015facial} show that the forehead and cheek regions yield the strongest rPPG signals. Only clear, complete facial views are processed further, utilizing an efficient active shape model (ASM)\cite{cootes1995active} for facial keypoint detection. This model controls shape distribution and predicts motion direction and position of keypoints. If a complete face isn't detected within 0.4s—equivalent to 10 video frames necessary for accurate rPPG detection—the system enters \textit{sleep} mode for 1s. Upon detecting complete faces in short-term mode, we analyze time-domain features of the rPPG signal, such as the pNN50 HRV index\cite{akselrod1981power,electrophysiology1996heart}, to identify critical events like \textit{sinus bradycardia} and \textit{high-degree atrioventricular block}.
% Studies, such as ~\cite{irani2014improved} and ~\cite{moreno2015facial}, have shown that the forehead and cheek regions of the face contain the strongest portion of the rPPG signal. 
% Frames with unclear or partial face views do not require further input.
% For facial keypoint detection, we utilize an efficient active shape model (ASM)~\cite{cootes1995active}. 
% ASM statistically models shape vectors to control a reasonable distribution, while a local gradient statistical model determines the motion direction and position of facial keypoints during prediction iterations.
% %
% If a complete face is not detected within the short-term sampling mode, the system switches to the \textit{sleep} mode. 
% The short-term sampling mode lasts for 0.4 seconds, equivalent to at least 10 consecutive video frames, found experimentally to be necessary for accurate rPPG detection. 
% If no valid facial image is captured, the system enters \textit{sleep} mode for 1 second.
% Upon continuous detection of complete facial images in short-term capture mode, we analyze time-domain features of the rPPG signal, such as the pNN50 heart rate variability (HRV) index~\cite{akselrod1981power,electrophysiology1996heart}, to identify key heart rate-related events. Through consultation with medical professionals, \sysname primarily identifies \textit{sinus bradycardia (slow heart rate)} and \textit{high-degree atrioventricular block (rapid heart rate)} events.
Specifically, pNN50 measures the percentage of adjacent heartbeats with intervals differing by more than 50ms. A value above 20\% indicates high heart rate variability, requiring at least 4 minutes of recording for reliable HRV assessment. In long-term sampling mode, if the pNN50 exceeds 20\%, sampling continues until it drops below 20\% or the user exits the camera's view. If pNN50 is under 20\%, indicating stability, the camera switches to sleep mode, alternating with short-term sampling for heart rate monitoring. If the average heart rate changes significantly (more than 5$\times$), the system resumes long-term sampling. During critical events, \sysname enters long-term mode to monitor continuous heart rate fluctuations, processing until 10 consecutive frames lack facial data.}

%但是，如果在长时间的检测中用户的心率都处于稳定的状态，不间断的长期捕获模式也会耗费大量不必要的能量。因此我们还需要根据用户实际的心率状态设计一种自适应间隔启动长期捕获模式的机制。在临床和生理学领域，有许多关于人体心率变化的研究。心率异变性(HRV)是衡量心率变化的一个重要指标[], pNN50则是衡量心率异变性的一个时域指标。pNN50是HRV分析中的一个指标，它表示相邻的两个心跳之间的时间间隔差异大于50ms的比例，高于20%被视为高心率异变性。[]指出至少需要4分钟的记录时间来获得可靠的HRV结果。因此，每当长期捕获模式持续4分钟，我们计算用户的平均心率与pNN50值，若低于20%，我们认为用户处于相对静止状态，系统睡眠模式与短期捕获模式的循环，并在短期捕获模式也进行心率感知，若平均心率变化超过5次，系统重新长期捕获模式。若高于20%，则继续长期捕获模式，直到pNN50值低于20%或用户离开相机镜头范围。

\begin{figure}[t]
    \centering
    \includegraphics[width=0.63\textwidth]{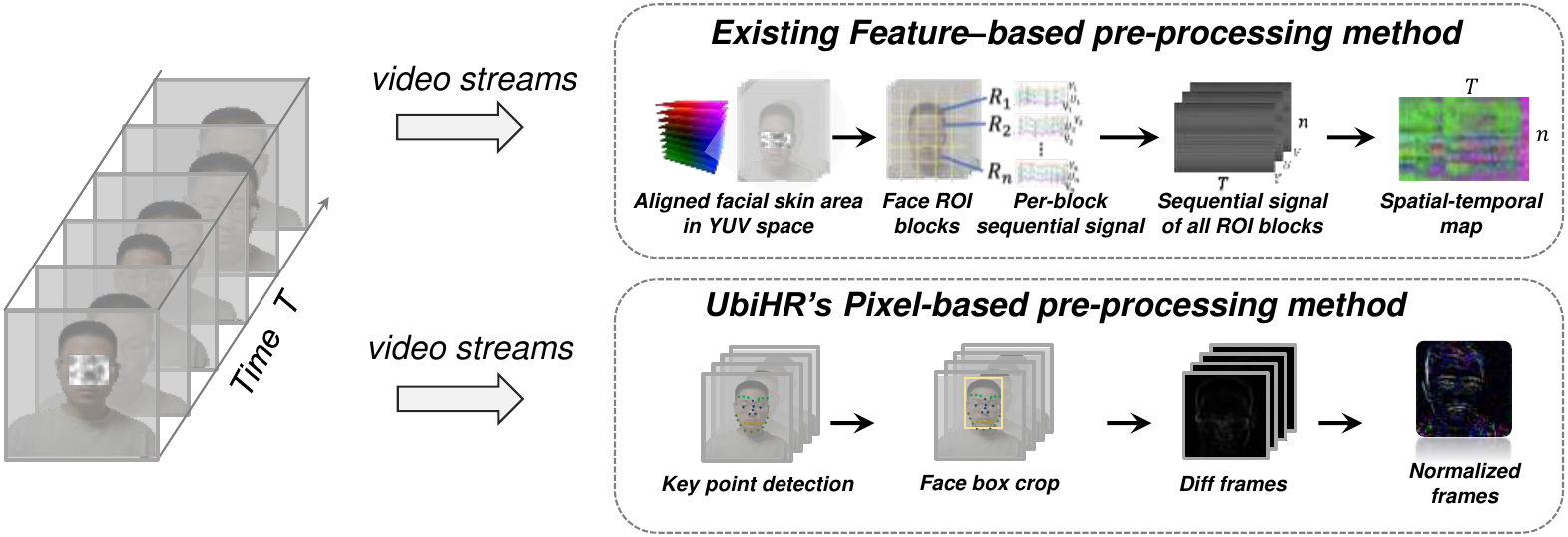}
    \caption{\lsc{Comparison of UbiHR and existing pre-processing methods.} }
    \vspace{-3mm}
    \label{fig:preprocess comparison}
\end{figure}

\subsection{\lsc{Dynamic Noise-aware Facial Image Pre-processing}}
\label{subsec:preprocess}
% \TODO{First highlight the "Mobile Noise", \ie  dynamic lighting condition and motion noises in mobile environment, and then present the challenges and primer work's limitations, and finally give our design}
% \TODO{After obtaining the signals related to HR, a common operation is to apply Fast Fourious Transform (FFT) to the signals to get the spectral distribution. The peak of the spectral distribution is considered as the HR frequency (in Hz). Directly applying FFT to the HR signals may get a spectral distribution consisting of many spectral noises. Therefore, a few approaches have been proposed to reduce such noises. In [21], Kumar et al. used the frequency characteristics as the weights to combine the HR signals from different ROIs in order to get a better HR signal. Instead of enhancing the HR signals, Wang et al. [22] directly adjusted the spectral distribution according to the prior knowledge of the HR signal. Similarly, Niu et al. [23] took the continuous estimation situations into consideration and used preceding estimations to learn an HR distribution and use it to modulate the spectral distribution. While the frequency domain representation is helpful for remote HR estimation, temporal domain information can also be helpful.
% All the methods mentioned above mainly focus on improving the hand-crafted pipelines from periodical temporal signal extraction to signal analysis. Their generalization ability can be poor under unconstrained scenarios. }

\re{User motion and open environment lighting introduce noise in video frames, necessitating preprocessing for heart rate sensing. 
Extracting the heart rate directly from raw videos yields a signal contaminated with various interferences. 
Additionally, the mobility of cameras, such as those on smartphones and robots, complicates matters. 
Mobile cameras can measure heart rates under any condition but introduce instability and varying angles, causing shaking and lighting changes, which add noise to the heart rate signal. 
Therefore, preprocessing is crucial to improve the signal-to-noise ratio of the heart rate signal. 
Prior methods reduce video noises or enhance signal-to-noise ratios for heart rate sensing, mainly include four categories:}
\textit{i)} Independent Component Analysis (ICA)~\cite{poh2010non,poh2010advancements}, Principal Component Analysis (PCA)\cite{lewandowskameasuring}, and constrained ICA (cICA)\cite{tsouri2012constrained} can improve the signal-to-noise ratio of the HR signal from RGB frame sequences.
\textit{ii)} Combining HR signals from different Regions of Interest (ROI) using frequency-based weighting~\cite{kumar2015distanceppg}.
\textit{iii)} Dividing the face into multiple ROI regions to obtain a temporal representation matrix and utilize matrix completion to purify HR-related signals.
\textit{iv)} Extracting the feature map from the green channel of frames~\cite{hsu2017deep}.

\re{However, they struggle to balance accuracy and latency (as we will evaluate in \secref{sec:evaluation}).
Mathematical and curve-based methods i), ii), and iii) mathematical and curve-based methods provide quick preprocessing but limited signal-to-noise improvements in heart rate signals. 
Method (iv), while promising, is time-consuming, with the STMap approach~\cite{niu2019rhythmnet} requiring complex calculations and over 200ms per frame to refine HR signals. 
To address these challenges, we introduce a novel preprocessing approach that quickly adapts to motion and lighting variations. It employs motion modeling and tailored normalization strategies across consecutive RGB frames, effectively reducing noise from these common real-world variables
}

\lsc{
In particular, we present a \textit{temporal difference layer}, a dedicated layer designed to compute motion disparities between adjacent frames during pre-processing before input to the deep learning model. 
Specifically, we employ temporal difference by evaluating brightness changes between adjacent frames, considering factors such as current modulation intensity, specular reflection, diffuse reflection, and optical sensor noise. 
This differential analysis enables the quantification of differences in diffuse reflection intensity induced by motion.
Referencing \equref{eq:option}, which illustrates the optical foundation of differential frames:
} 
\begin{equation}
    C_{k}(t) = (I(t) \cdot (v_{s}(t) + v_{d}(t)) + v_{n}(t)) - (I(t-1) \cdot (v_{s}(t-1) + v_{d}(t-1)) + v_{n}(t-1)) \label{eq:option}
\end{equation}
\lsc{where $C_{k}(t)$ represents every two consecutive frames and $I(t)$ denotes the modulated luminance intensity influenced by specular reflection $v_{s}(t)$, diffuse reflection $v_{d}(t)$, and optical sensor quantization noise $v_{s}(t)$.}

\lsc{
Furthermore, to address the challenge of significant variations in signal scale across frames, especially when the heart rate signal of interest is obscured by subtle pixel variations along the temporal axis, and noise artifacts can result in significantly larger relative changes., we introduce a \textit{learnable batch normalization} layer~\cite{liu2023efficientphys} after the temporal difference layer. 
\re{This normalization layer incorporates \textit{two learnable parameters}, $\beta$ and $\gamma$, scaling to different variances and shifting to dynamic signal means.}
This normalization layer ensures that the difference frames within a batch are normalized to the same scale. 
\equref{eq:batchnorm} demonstrates how the batch normalization layer learns optimal parameters that amplify pixel changes while simultaneously minimizing the impact of noise. 
}
\begin{equation}
    N_k(t) = \frac{(\beta_t * D_k(t) + \gamma_t)-\mu D_k}{\sigma D_k} \label{eq:batchnorm}
\end{equation}
 Where \begin{math}\mu\end{math} is mean and \begin{math}\sigma\end{math} is standard deviation.
\lsc{
Without this normalization step, the frame differences tend to be close to zero, given that subtle variations in user facial skin pixel values between consecutive frames are relatively small. 
Consequently, batch normalization not only amplifies numerical differences but also normalizes them. 
Figure \ref{fig:preprocess comparison} compares the contrasting pre-processing workflows of existing methods and ours.
}

\begin{figure}[t]
    \centering
    \includegraphics[width=0.7\textwidth]{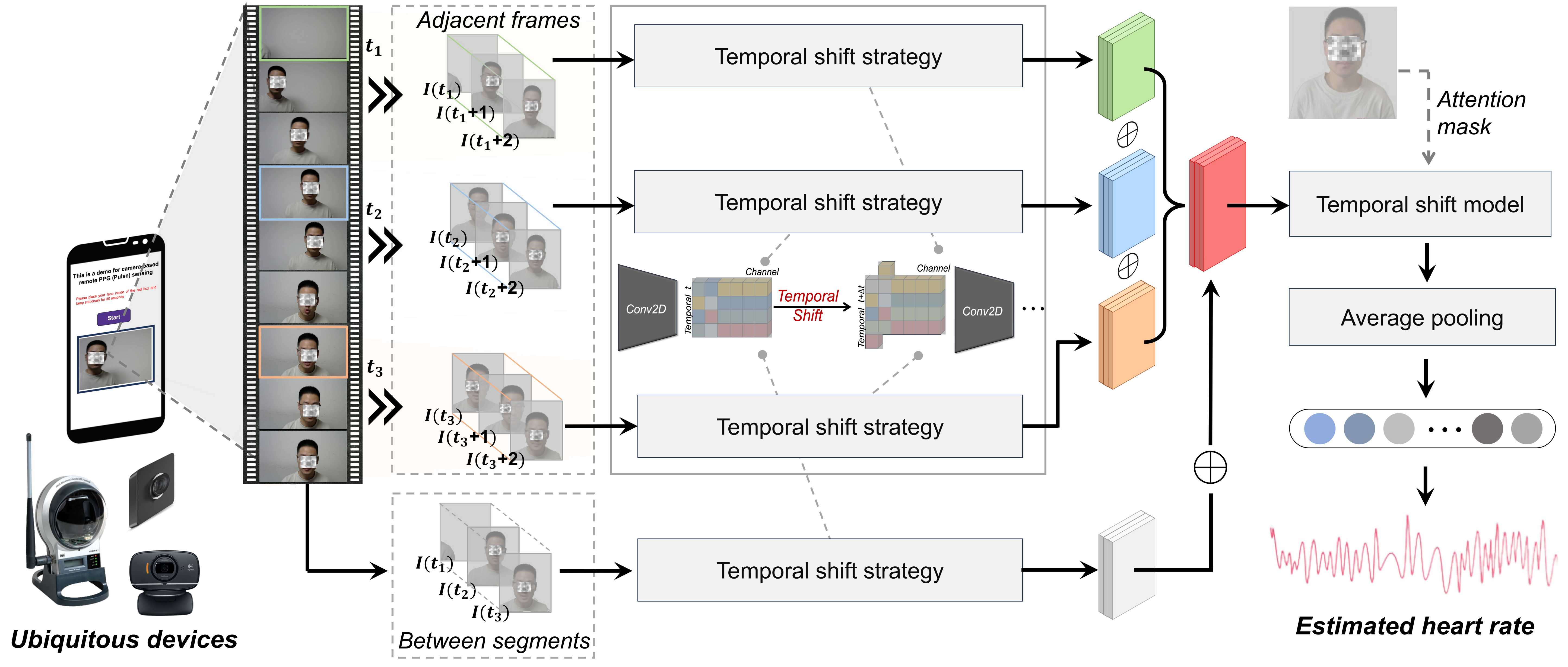}
    \vspace{-3mm}
    \caption{\lsc{The architecture of the heart rate recognition model (the mosaic in this paradigm is to protect privacy)}.}
    \label{fig:architecture}
\end{figure}

\subsection{\lsc{Fast Long-range Spatio-temporal Heart Rate Recognizing}}\label{sec:3.4}

\subsubsection{Primer on Long-range Spatio-temporal Model for HR Detection}\label{sec:3.4.1}
%与视频理解任务相似，基于摄像头的远程心率检测也可以认为是从视频序列到信号序列的映射问题，即视频帧中空间维度光线反射与心率信号在时间维度上变化的映射关系。由于心率信号本身的特点，如图8所示，无论用户的心跳是快还是慢，心跳幅度大还是小，来自不同时间位置的心率信号在波峰位置都具有相似的特性：信号趋势的突变以及相对高的信号强度，因此在long-range范围构建时空模型能够使得检测更精确与鲁棒。此外，如上文所提出的实际应用场景，camera-based HR detection通常是一项长时间的监控任务，这也需要模型具备长时间的时空感知能力。

\begin{figure}[t]
    \centering
    \includegraphics[width=0.3\textwidth]{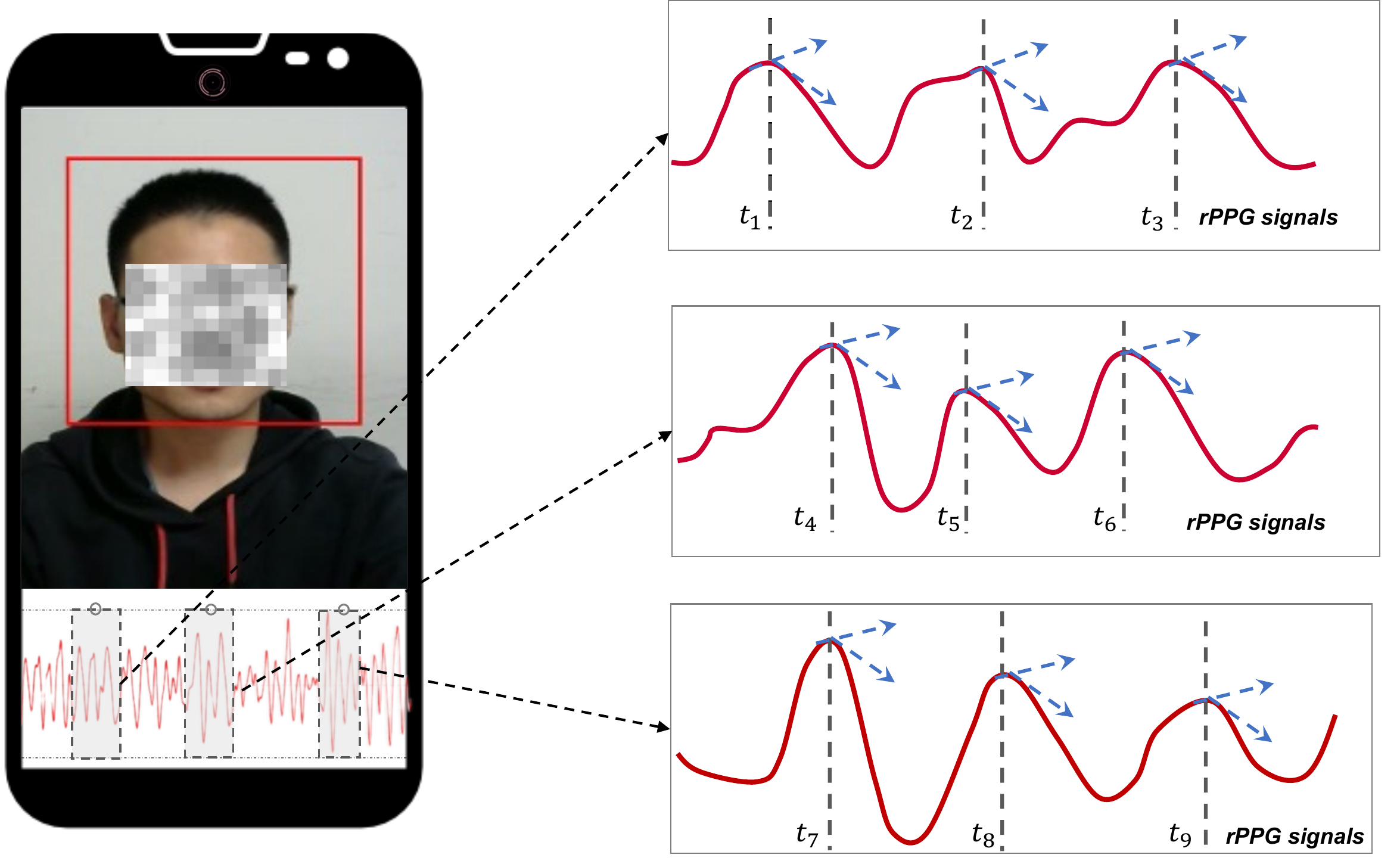}
    \caption{\lsc{The user's heart rate wave varies, while the abrupt transitions at the peak rPPG signal are consistent.}}
    \label{fig:long-term}
\end{figure}

\lsc{Camera-based heart rate recognition involves mapping video sequences to signal sequences. 
Specifically, it translates the spatial dimension of light reflection in video frames into temporal variations of heart rate signals.
As depicted in Figure \ref{fig:long-term}, peak positions exhibit similar characteristics regardless of diverse user heart rate speed or amplitude. 
Our key observation is that these characteristics include abrupt changes in signal trends and relatively high signal intensity. 
Therefore, constructing a long-range spatio-temporal model can enhance detection accuracy and robustness.
}

\begin{itemize}
    \item \textbf{\lsc{Limitations of Short-range Methods.}}
    \lsc{Most prior methods extract spatio-temporal rPPG features from \textit{short ranges}~\cite{bousefsaf20193d,yu2020autohr,vspetlik2018visual}, such as adjacent frames, overlooking long-range relationships among periodic rPPG features, thereby compromising stability against noise.
    As illustrated in \tabref{tab:noise}, compared to scenarios with slight head movements (Task 1), both PhysNet \cite{yu2019remote} and DeepPhys \cite{chen2018deepphys} experience significant Mean Absolute Error (MAE) increases of 6.34 and 10.71, respectively, in high-noise scenarios, such as when the user is speaking (Task 2).
    Additionally, as mentioned in \secref{subsec:principle}, in practical applications, users often require long-range spatio-temporal awareness of heart rates.}
    
    \item \textbf{\lsc{Limitations of Long-range Physformer.}}
    \lsc{
    A typical long-range spatio-temporal model, \ie Physformer \cite{yu2022physformer}, introduced a transformer architecture that yielded promising accuracy. 
    However, it has 7.03M parameters and 47.01GFLOPs, cannot fit into resource-limited mobile and embedded devices.}

    % \item \textbf{\lsc{Our Scope}}. 

\end{itemize}

%但是，大多数以前的方法仅考虑来自相邻帧的时空RPPG特征，而忽略了准周期性RPPG特征之间的远距离关系。这导致他们抗噪声的能力弱。如表所示，与轻微的头部运动场景相比，PhysNet和DeepPhys在用户说话这种高噪声场景下，平均错误率(MAE)分别上升了%和%。 也有一些研究致力于long-range时空建模，Physformer引入transformer架构实现了较好的结果，但是它具有7.03M和47.01GFLOPs，处理资源开销大，对移动端设备部署不友好。因此，我们提出轻量化的新模型设计。
 
% 1）已有的抗噪声干扰的能力弱，因为是基于相邻帧的xxx模式。 【给出1-2个数据，例如，在xxx噪声下【罗列】,xxx方法精度下降多少 -小实验表格】【并给出原理解释】-----》 2） 提出我们的long-term 长时间序列的模型，为了抗噪 【参考文献--长时序的抗噪能力更强，原理 --- 文章找到发我】 
% 但是，这种长时间序列模型非常大，处理资源开销很大。因此，我们提出轻量化的新模型设计。 1） 3D CNN在心率监测中的优势是 xxx 1) 2) 3) ？ " 用2D CNN 和TSM 实现3D CNN的效果" 。 2) 多分枝并行是为了长时序抗噪，原理，价值。 快，轻量。  3）引入TSM还以引入额外的噪声，看文献-改词。修正噪声。我们的原理是什么：soft-mask。 额外什么噪声怎么引入的，引入的什么，如何影响了精度的什么--原理 。

\begin{table}[]
\centering
\scriptsize
\caption{Error of two existing short-range methods with different noises.}
\vspace{-3mm}
\begin{tabular}{|c|cc|}
\hline
\multirow{2}{*}{\textbf{Short-range Methods}} & \multicolumn{2}{c|}{\textbf{MAE}}                    \\ \cline{2-3} 
                                 & \multicolumn{1}{c|}{\textbf{Task1 (Slight head motion)}} & \textbf{Task2 (Noisy, user is speaking)} \\ \hline
\textbf{PhysNet~\cite{yu2019remote}}                 & \multicolumn{1}{c|}{1.86}           & 8.20           \\ \hline
\textbf{DeepPhys~\cite{chen2018deepphys}}                & \multicolumn{1}{c|}{2.11}           & 12.82          \\ \hline
\end{tabular}
\label{tab:noise}
\end{table}

    \re{
    To address these, \sysname need: \textit{i)} operate in real-time to learn noise-robust representations corresponding to pulse and respiration signals from long-range spatio-temporal features, }\textit{ii)} be low-cost on resource-constrained mobile and embedded devices,
    Specifically, even with improved signal-to-noise ratio through preprocessing (see \secref{subsec:preprocess}), performing rPPG-based heart rate measurement remains challenging due to the subtle amplitude of light absorption variation \textit{inherent} in heart rate signal itself~\cite{lu2021dual}. 
    For example, color distortion, pseudo-color, or color noise caused by camera signal processing algorithms.
    Also, rPPG signals are sensitive to interference from head motion, lighting conditions, and sensor noises, complicating the construction of a robust tool in less-constrained ubiquitous environments. 
    Therefore, we present a parallel multi-branch model to extract spatio-temporal features across both short and long ranges (\secref{subsec:multibranch}), with a lightweight spatio-temporal model (\secref{sec:2DCNN}).

\subsubsection{Incorporating Temporal Shift to 2D Convolution for Lightweight Spatio-temporal Modeling}
\label{sec:2DCNN}

\begin{figure}[t]
\centering %图片全局居中
%并排几个图，就要写几个minipage
\includegraphics[width=0.8\textwidth]{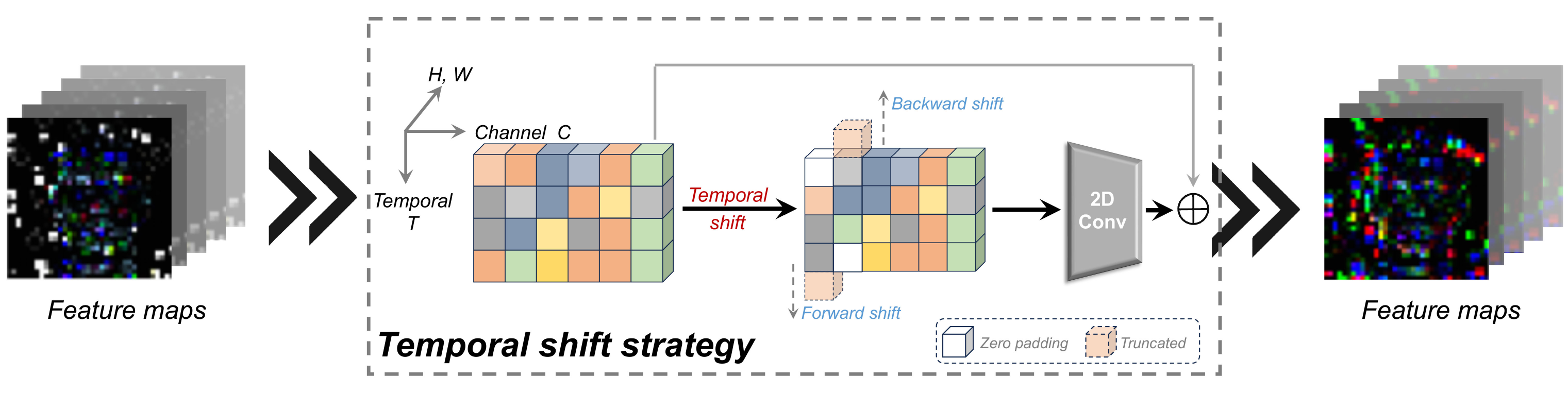} 
\caption{Illustration of incorporating temporal shift to 2D convolution.}
\label{fig:TSM}
\end{figure}

\re{We utilize a temporal shift strategy to enhance the temporal modeling of 2D convolutions, offering noise-robustness and awareness similar to 3D convolutions, but without the extra computational burden. 
\textit{3D convolutions} \cite{yu2019remote} intuitively enhance the model's robustness and awareness by capturing spatial and temporal variations in frame sequences, effectively mapping heart rate signals. 
These convolutions progressively extract finer features through multiple layers, differentiating small heart-rate-related signals from larger noise. 
Additionally, we chose convolutional models over RNN or LSTM because they allow for parallel processing, share kernels across dimensions to reduce parameters and avoid the gradient issues common in sequence processing. 
Yet, the computational demands of 3D convolutions challenge real-time performance on limited-resource devices.
}

\lsc{ 
Therefore, we integrate the zero-parameter temporal shift strategy into 2D Convolution. 
The temporal shift strategy depicted in Figure \ref{fig:TSM} enhances temporal modeling efficiency by shifting feature maps along the temporal dimension. 
It partitions the input tensor into three blocks along the channel dimension, shifting the first block leftward (advancing time by one frame) and the second block rightward (delaying time by one frame). 
\re{These shifts enrich spatial convolution with temporal feature variations.} 
Notably, tensor shifting introduces no additional parameters to the DL model but facilitates information exchange between adjacent frames.
}
As we will show in \secref{sec:evaluation}, the lightweight spatio-temoral model costs 49\% in memory and 3\% in latency lower for a 30 frames input.

\subsubsection{Multi-branch for Long- and Short-range Spatio-temporal Modeling}
\label{subsec:multibranch}
\lsc{Despite its utility, the temporal shift strategy remains confined to short-range spatio-temporal information, limiting its effectiveness in capturing long-range periodic variations in light reflection from sequential video frames. 
This limitation blocks the model's ability to effectively map long-range heart rate signals.
As mentioned in \secref{sec:3.4.1}, capturing the heart rate signal's periodic characteristics over the long-range is pivotal for suppressing non-physiological information like noise, thus enhancing detection accuracy and robustness.
}

\lsc{
To overcome this challenge, we introduce a multi-branch parallel spatio-temporal model capable of both short- and long-range spatio-temporal modeling. 
As illustrated in \figref{fig:architecture}, we partition the input frame sequence into three groups and apply a single-branch temporal shift+2D convolution operation on three adjacent frames, \ie short-range, within each group. 
These \textit{short-range branches} function in parallel, combining their results to yield short-range outcomes similar to prior approaches, referred to as the \re{\textit{"adjacent frames branch"}}.
Concurrently, we select one frame from each group in the \re{\textit{adjacent frames branch}} to generate a new frame sequence. 
We conduct a single-branch temporal+2D CNN convolution operation independently on this sequence, creating the "\textit{segment branch}." 
Operating between frames with longer temporal distances, the segment branch facilitates long-range spatio-temporal modeling.
Finally, we merge the outputs of the \textit{adjacent frame branch} (for short range) and the \textit{segment branch} (for long range), performing convolution operations to capture heart rate features consistent across both branches. 
Accurate mixed-granularity results are then obtained after average pooling. 
}

\subsubsection{Towards Parallel Speedup}
\lsc{
The multi-branch parallel spatio-temporal model enhances computational efficiency by enabling concurrent convolution operations in the adjacent frames branch, eliminating the sequential processing bottleneck. 
This parallel execution ensures that subsequent frames can be processed independently of each other, significantly reducing waiting time. 
Additionally, implementing asynchronous processing further enhances parallelism, reducing waiting time. 
By grouping consecutive video frames into groups, we exploit parallel capabilities more effectively, accelerating the overall process.
}

\subsubsection{Soft-attention Mask-based Accuracy Recorrecting}
\label{sec:attention}
%TSM在时间维度上移动tensor会将其移动的所有内容都融合到相邻帧中并替换原来帧中的信息，即使是噪声。这可能会使HR信号强的部分被替换为噪声信息，或是在本应是噪声的区域引入当前帧中不存在的HR信号信息。
\lsc{Incorporating the temporal shift strategy introduces additional temporal information, potentially leading to noise interference in the processed frames in reverse~\cite{liu2020multi}.
This is because temporal shift integrates all the content it moves into the adjacent frames and replaces the original information in those frames, including noise. 
This can result in the replacement of high-intensity parts of the heat rate signal with noise information or the introduction of heart rate signal information in regions where noise should exist in the current frame. 
Therefore, it is crucial to prioritize pixels that already contain heat rate information over those affected by temporal shift-induced noises.}

\lsc{To mitigate this, we propose the integration of \re{\textit{soft-attention masks} after each branch and before average pooling. }
These masks prioritize pixels with existing heart rate information over those affected by temporal shift-induced noise, allowing the network to focus on target signals.
Soft-attention masks prioritize pixels that display stronger signals in intermediate convolution representations by assigning them higher weights, reducing the impact of noise. 
Generated through $1\times 1$ convolutions with pre-processed frames as input, these masks are computed using the sigmoid activation function. 
Element-wise multiplication is performed with the respective representations from the merge outputs of the adjacent frames branch and segment branch. 
This strategy ensures that the network effectively discerns and prioritizes pixels with meaningful HR information while attenuating noise interference from the temporal shift.
The computation of the attention mask is shown in \equref{eq:mask}:}
\begin{equation}
     \frac{H_{j}W_{j} \cdot \sigma(\omega^{j} 
     \mathbb{X}_{\alpha}^{j} + b^j) }{2 \parallel \sigma(\omega^j \mathbb{X}_{\alpha}^j + b^j) \parallel_1} \label{eq:mask}
\end{equation}
\lsc{where $j$ is the index of the convolution layer, $\omega^j$ is a 1x1 convolution following the sigmoid activation function $\sigma()$. 
And $\mathbb{X}_{\alpha}^{j}$ is the feature maps in the convolution layer, $H_j$ and $W_j$ is the height and width. $b_j$ is the bias. 
Finally, we perform element-wise multiplication with the respective representation from merge outputs of adjacent frames branch and segment branch.
}
\re{As a separate note, if detection remains incorrect, we re-inference erroneous frames and using a longer sequence of video frames to refine the results}.

%% file: body/5-evaluation.tex
\section{EVALUATION}\label{sec:evaluation}
\lsc{This section presents the experimental settings and system performance of \sysname.}
% to comprehensively evaluate the performance of .

\subsection{Implementation}
\lsc{We train the DL model on the server while conducting heart rate video sampling, detection, and recognition under various motion and lighting conditions on mobile and embedded devices.
On the \textit{server side}, \sysname is implemented in PyTorch (v1.12.1) and utilizes the Adadelta optimizer with a learning rate of 1.0, a batch size of 32, and a 3x3 kernel size with 2x2 pooling. 
We apply dropout rates of 0.25 and 0.5. 
All spatio-temporal models use a window size of 10 frames for fair comparison. 
Model training is accelerated using Nvidia TITAN RTX with CUDA 10.2. 
On the \textit{client side}, we package the system as a Python module for seamless integration with other applications. 
Video sampling is performed at 30fps.
We integrate existing methods into platforms like Android, primarily overcoming the limited availability of DL-based face recognition libraries.
The user interface comprises a viewfinder and a heart rate result display. 
We design platform-specific user interfaces for various mobile and embedded platforms. 
The camera viewfinder shows the live feed captured by the camera, with a red box indicating the actual framing area in the center. 
\re{To protect user privacy, all video frames captured during the duty-cycling sampling process are immediately processed and deleted locally, not sent to the cloud for preprocessing or retained for future processing. }
Results are displayed as waves using JavaScript, and hovering over the results reveals specific information about corresponding points.
}

\subsection{Experiment setups}

\textbf{Dataset and tasks}.
\lsc{We experiment with four datasets.
First, both UBFC-RPPG~\cite{bobbia2019unsupervised} and UBFC-Phys~\cite{sabour2021ubfc} are recorded by low-end cameras instead of professional high-end cameras, under varying sunlight and indoor illumination.
UBFC-RPPG features stationary subjects, while UBFC-Phys includes diverse head movements, introducing motion noise. 
We train on UBFC-RPPG and test on UBFC-Phys to evaluate system performance under diverse motion and lighting noises.
Second, we conduct tests on two synthetic datasets, SCAMPS~\cite{mcduff2022scamps} and UCLA-rPPG~\cite{wang2022synthetic}. 
SCAMPS synthesized 2800 RGB videos and provided corresponding ground truth PPG signals.
It also randomly rendered hair, clothing, and the environment to simulate various real-world scenarios. 
UCLA-rPPG synthesized 480 videos based on real videos and PPG signals. 
It rendered different actions, skin tones, and lighting conditions.
Specifically, We first train a backbone with four convolutional layers followed by a tanh activation function for input sequence using UBFC-rPPG.
It is captured from a low-end camera (Logitech C920 HD Pro) operating at 30fps and a resolution of 640x480 in uncompressed 8-bit RGB format. 
Ground truth PPG data, including waveform and heart rates, was acquired using a CMS50E transmissive pulse oximeter. Subjects were positioned approximately 1 meter from the camera, facing it directly.
We then employ UBFC-Phys to test system performance. 
UBFC-Phys are collected from 56 participants undergoing a three-step stress-inducing experience, \eg a relaxation task (T1), a speech task (T2), and an arithmetic task (T3), to dynamically affect the heart rates.
}

\textbf{Ubiquitous devices}.
\lsc{We test \sysname with four commercial mobile camera-embedded platforms, including:
\textit{i)} Device $D_1$: Raspberry Pi 4B, equipped with a 4-core 1.5GHz ARM Cortex-A72 CPU, \re{is a low-power, single-board computer that is widely used in various embedded and edge computing applications};
\textit{ii)} Device $D_2$: Nvidia Jetson Orin, equipped with a 4-core ARM Cortex-A78AE CPU, \re{offers significantly higher compute performance compared to the Raspberry Pi, with up to 57 TOPS of AI inference capacity};
\textit{iii)} Device $D_3$: Nvidia Jetson AGX, equipped with an 8-core ARM Cortex-A78AE CPU, \re{is a high-performance AI computing device designed for advanced robotic, autonomous systems, and edge AI applications}
and iv) Device $D_4$: IQOO9 smartphone with the Snapdragon 8Gen1 8-core CPU.
% These devices are ubiquitous platforms for small robots, cameras, \etc
With any embedded device equipped with a camera, \sysname can transform it into a 'heart rate sensing expert assistant,' offering health services in users' daily lives at home.
}

\textbf{Baselines}. \lsc{We compare the performance of \sysname with the following camera-based HR detection baselines.}
\begin{itemize}
    \item \textbf{Mathematical/shallow machine learning-based physiological detection methods:} 
    \begin{itemize}
        \item \textbf{GREEN}~\cite{verkruysse2008remote}: It utilizes the optical reflection of green spectrum on skin blood flow for measurement, which is simple to implement but susceptible to environmental light interference, and is a feature-based method relying on skin optical reflection characteristics.
        \item \textbf{CHROM}~\cite{de2013robust}: It employs a linear combination of chrominance signals, assuming a standardized skin color profile. 
        \item \textbf{POS}~\cite{wang2016algorithmic}: It separates the heart rate information from the optical signal using the principle of pulse occlusion, which can suppress respiration and motion interference, relying on the pulse occlusion model.
        \item \textbf{LGI}~\cite{pilz2018local}: It investigates the influence of prior knowledge regarding invariance on heart rate estimation from face videos captured in real-world conditions. 
        \item \textbf{PBV}~\cite{de2014improved}: It obtains the heart rate information by measuring changes in local skin blood volume, relying on the photoplethysmography blood volume model.
        \item \textbf{ICA}~\cite{poh2010advancements}:  It separates the heart rate component from multi-channel optical signals, and is relatively robust to motion noises.
    \end{itemize}
    
    \item \textbf{DL-based HR detection methods}:
    \begin{itemize}
        \item \textbf{DeepPhys}~\cite{chen2018deepphys}: 
        \lsc{It employs a deep convolutional network for video-based heart and breathing rate sensing.} 
        \item \textbf{PhysNet}~\cite{yu2019remote}: it uses a deep spatio-temporal network for reconstructing rPPG signals from facial videos.
        \item \textbf{On-device latency-efficient TS-CAN}~\cite{liu2020multi}: It leverages a temporal shift convolutional attention network (TS-CAN) based on hand-crafted pre-processing, to enable HR detection on mobile platforms.
        \item \textbf{On-device latency\&memory-efficient \sysname}: \lsc{It enables on-device, real-time, and low-memory HR detection through efficient pixel-based preprocessing and a multi-branch parallel spatio-temporal network.}
    \end{itemize}

\end{itemize}

\lsc{
\textbf{Evaluation metrics}.
We employ three categories with five metrics to evaluate accuracy as heart rate signals dynamically change over time.
\textit{First}, we assess \textit{absolute differences} between predicted and true values using \textit{Mean Absolute Error} (MAE) and \textit{Mean Absolute Percentage Error} (MAPE). 
While MAE measures absolute differences, MAPE offers a percentage-based explanation, making error interpretation more intuitive.
\textit{Second}, we examine the \textit{relative relationship} and \textit{continuity} of continuous HR signals using \textit{root mean square error} (RMSE) and \textit{Pearson coefficient} to capture relative changes and temporal relationships. 
RMSE, akin to the L2 norm, is sensitive to outliers, unlike MAE, which represents the L1 norm.
\textit{Third}, we evaluate \textit{pulse signal-to-noise ratios} (SNR) to gauge system noise robustness. 
Ground truth HR frequency is obtained from contact pulse oximeters.
\textit{Also}, we test \sysname's latency and memory usage in real-world mobile and embedded systems.
}

\begin{figure}[t]
\centering  %居中
\subfigure[Device $D_1$: Raspberry pi]{   %第一张子图
\begin{minipage}{3.8cm}
\centering    %子图居中
\includegraphics[scale=0.2]{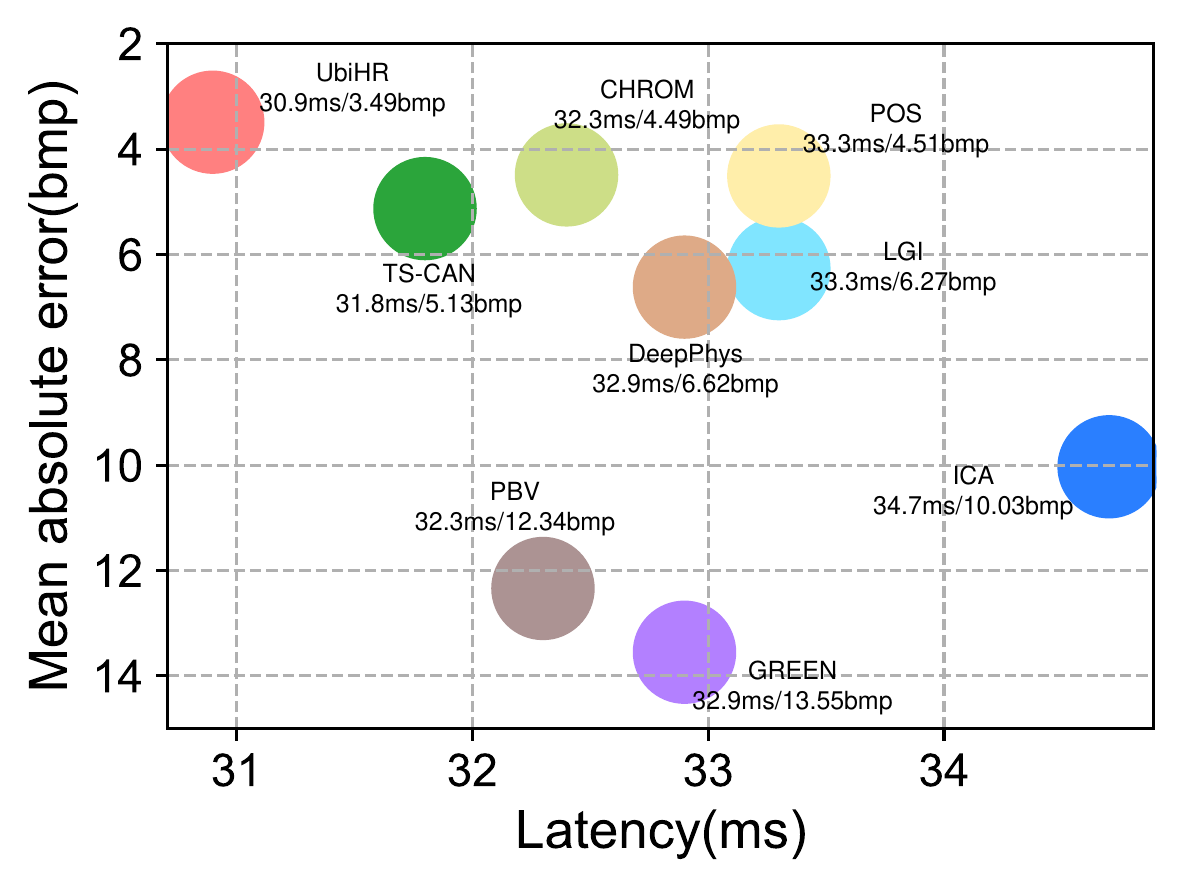}
\end{minipage}
}
\hspace{8mm}
\subfigure[Device $D_2$: Jetson Orin]{ %第二张子图
\begin{minipage}{3.8cm}
\centering    %子图居中
\includegraphics[scale=0.2]{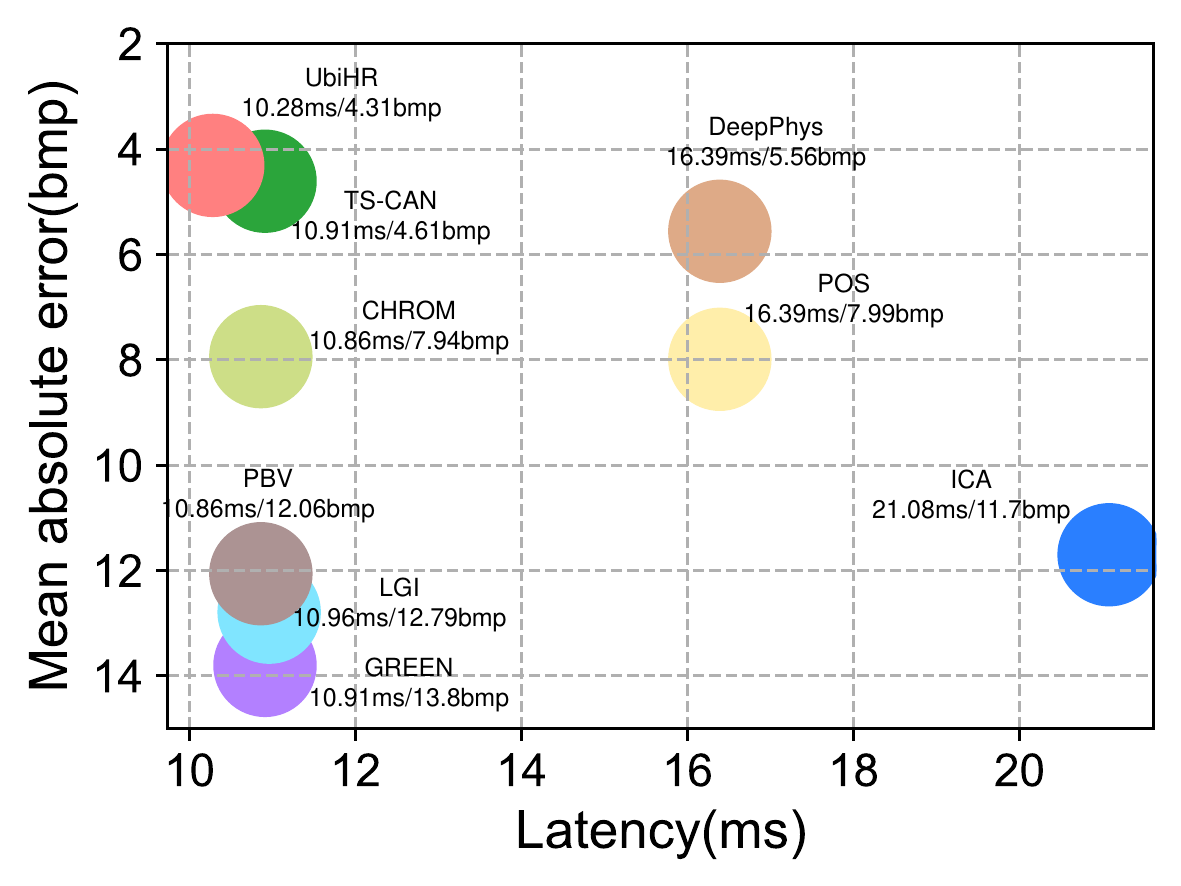}
\end{minipage}
}
\hspace{8mm}
\subfigure[Device $D_3$: Jetson AGX]{ %第三张子图
\begin{minipage}{3.8cm}
\centering    %子图居中
\includegraphics[scale=0.2]{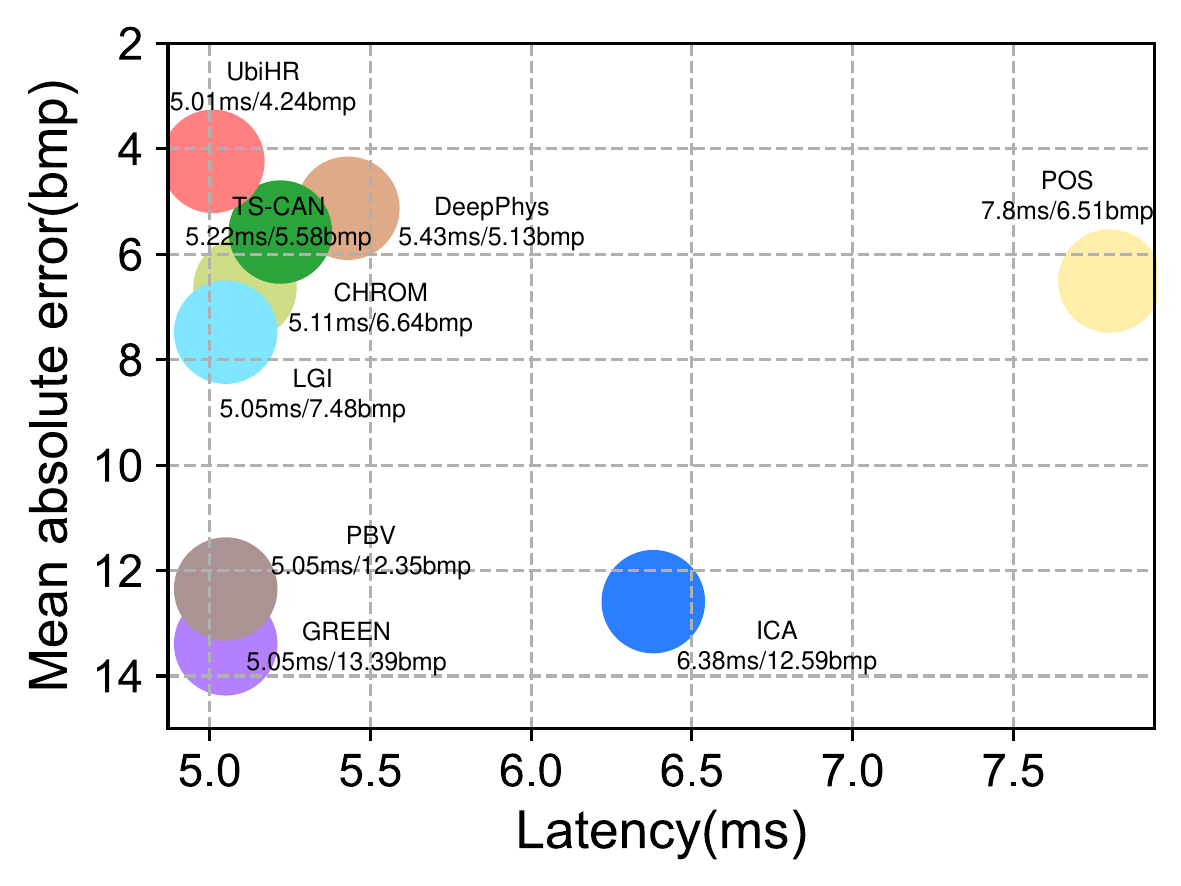}
\end{minipage}
}
\vspace{-3mm}
\caption{Trade-off comparison of eight baselines and \sysname.
\sysname achieves the best overall performance.}
\vspace{-3mm}
%大图名称
\label{fig:trade-off}    %图片引用标记
\end{figure}

\subsection{Performance comparison}
\lsc{This section presents the results of heart rate sensing compared with the baselines.
}
\subsubsection{Overall trade-off comparison}
\lsc{This experiment compares the performance of \sysname and nine different baseline methods on open video-based heart rate sensing datasets.
}

\textbf{Setups.}
%我们在以下三种设备上进行测试：设备一：树莓派4B，搭载4核心1.5GHz频率的ARM Cortex-A72 CPU。设备二：Nvidia Jetson Orin， 搭载4核心ARM Cortex-A78AE CPU。 设备三：Nvidia Jetson AGX， 搭载8核心Nvidia Jetson AGX。 三种设备的计算能力逐渐提高。 我们使用UBFC-Phys中部分不同任务的测试数据进行测试，记录并计算不同方法的时延。我们在每个设备上测试3次并分别取平均值。 由于DeepPhys使用的预处理方法无法在端上实现，我们仅测试了其only inerence的性能。
\lsc{We conduct tests on three typical mobile and embedded devices, \ie Device $D_1$ Raspberry Pi 4B, Device $D_2$ Nvidia Jetson Orin, and Device $D_3$ Nvidia Jetson AGX. 
We use test data of various tasks in UBFC-Phys, \ie participants performing three tasks under diverse natural lighting conditions: head motion (Task 1), speaking (Task 2), and diverse facial expressions (Task 3), to compare the accuracy, latency, and trade-off of eight different baselines and \sysname. 
We test three times on each device and calculate the average values.
} \re{we use the same inputs on each device, however, these three different devices have different CPU, and their resource constraints led to the use of different processing techniques and different third-party libraries on them.}

\textbf{Results.}
%UbiHR在三种测试设备上都取得了最好的成果
%function based方法普遍在时延上取得了较好的结果，这些基于手工制作模型的方法在无约束的数据上出现了难以接受的错误率。
%DeepPhys这种为了on sever设计的深度学习模型在不仅在检测时延上的成绩较差，需要复杂运算的模型在算力不足的设备上精度也出现了下降。
%作为为了on-device部署的TS-CAN在检测精度上取得了较好的结果，但是其预处理过程所耗费的时间在其总时延中占了较大比重。
Figure \ref{fig:trade-off} shows the results, indicating that \sysname outperforms other baselines across all testing devices. 
\textit{First}, curve-based methods, \eg GREEN, CHROM, ICA, LGI, PBV and POS, demonstrate decent latency performance, while these manually designed model-based methods exhibit unacceptable sensing errors on diverse noisy data. 
\textit{Second}, DeepPhys, tailored for server-side deployment, suffers from 59.4\% higher latency (only inference without pre-processing as its pre-processing is too complex to run on mobile and embedded devices) and reduced accuracy on resource-limited mobile and embedded devices due to high-cost pre-processing methods. 
We note that, due to the high-cost pre-processing methods used in DeepPhys being impractical to deploy on mobile and embedded devices, Figure \ref{fig:trade-off} only presents its inference latency without pre-processing. 
Even TS-CAN, designed for on-device deployment, processes over 10 frames slower per second than \sysname due to significant pre-processing and inference overhead.

\subsubsection{Accuracy comparison} 
\lsc{This experiment compares the heart rate sensing accuracy of baselines and \sysname.}

\lsc{
\textbf{Setups.}
We conduct evaluations across three tasks (Task 1 to 3) in UBFC-Phys on Device $D_1$ Raspberry Pi 4B.
Testing two standard accuracy metrics, Mean Absolute Error (MAE) and Mean Absolute Percentage Error (MAPE), with five measurements conducted and the average performance reported. 
} \re{The inputs used for testing different methods are the same.}

\begin{figure}[t]
    \centering
    \includegraphics[width=0.5\textwidth]{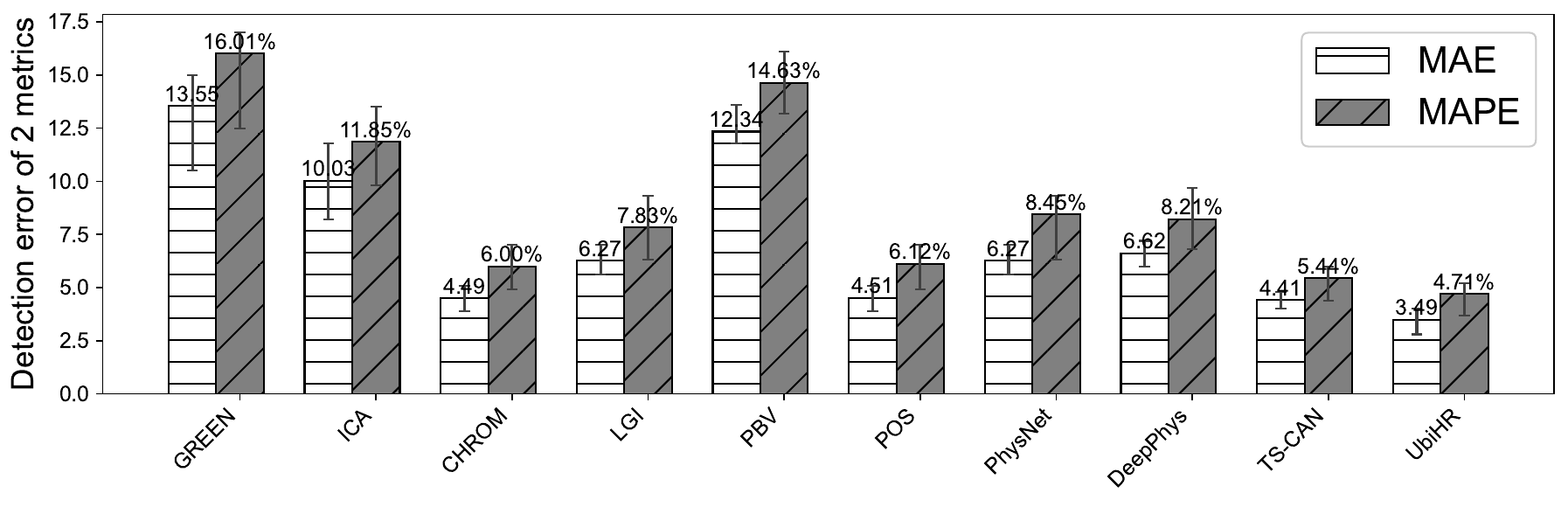}
    \vspace{-3mm}
    \caption{Detection error comparison of nine baselines and \sysname.}
    \vspace{-3mm}
    \label{fig:error_comparison}
\end{figure}

\begin{figure}[t]
\centering  %居中
\subfigure[\sysname]{   %第一张子图
\begin{minipage}{2.5cm}
\centering    %子图居中
\includegraphics[scale=0.25]{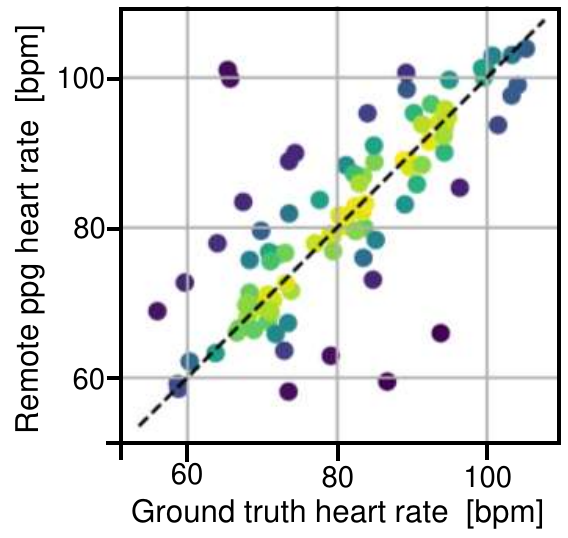}
\label{fig:bland-my}
\end{minipage}
}
\hspace{3mm}
\subfigure[DeepPhys]{ %第二张子图
\begin{minipage}{2.5cm}
\centering    %子图居中
\includegraphics[scale=0.25]{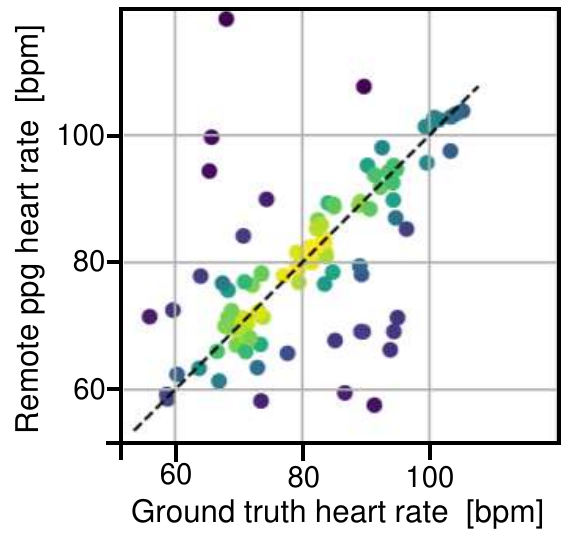}
\label{fig:bland-deep}
\end{minipage}
}
\vspace{-3mm}
\caption{\lsc{Comparison of DeepPhys and \sysname in fitting estimated heart rate to ground truth.}} 
\vspace{-3mm}
\label{fig:bland}    %图片引用标记
\end{figure}

\lsc{
\textbf{Results}.
\figref{fig:error_comparison} summarize the results.
\textit{First}, \sysname achieves the highest accuracy, significantly outperforming curve-based POS with a 22.6\% reduction in Mean Absolute Error (MAE) and a 1.32\% decrease in Mean Absolute Percentage Error (MAPE).
This superiority is because of the nuanced nature of heart rate variability features, where manually crafted methods like POS may struggle with consistency and are vulnerable to noise.
\textit{Second}, compared to existing DL-based methods like TS-CAN, \sysname reduces MAE by 32\% and MAPE by 0.73\%, indicating its improved capability to capture intrinsic heart rate features and robustness against noise.
\textit{Third}, a Bland-Altman plot (Figure \ref{fig:bland}) visualizes the correlation between \sysname's and DeepPhys' predicted results and ground truth.
Comparing the plots, \sysname demonstrates a stronger correlation between predicted and actual heart rates, particularly in the 60 bpm to 110 bpm range of the UBFC-Phys dataset, with fewer outlier values.
}

% \begin{figure}[h]
%     \centering
%     \includegraphics[width=0.45\textwidth]{./pictures/example.png}
%     \caption{example}
%     \label{fig:example}
% \end{figure}

% \begin{figure}[H]
% \centering %图片全局居中
% %并排几个图，就要写几个minipage
% \begin{minipage}[b]{0.45\textwidth} %所有minipage宽度之和要小于1，否则会自动变成竖排
% \centering %图片局部居中
% \includegraphics[width=0.9\textwidth]{./pictures/physnet.png} %此时的图片宽度比例是相对于这个minipage的，不是全局
% \caption{}
% \label{fig:physnet}
% \end{minipage}
% \begin{minipage}[b]{0.45\textwidth} %所有minipage宽度之和要小于1，否则会自动变成竖排
% \centering %图片局部居中
% \includegraphics[width=0.9\textwidth]{./pictures/deepphys.png}%此时的图片宽度比例是相对于这个minipage的，不是全局
% \caption{}
% \label{fig:deepphys}
% \end{minipage}
% \end{figure}

% \begin{figure}[h]
% \centering %图片全局居中
% %并排几个图，就要写几个minipage
% \begin{minipage}[b]{0.3\textwidth} %所有minipage宽度之和要小于1，否则会自动变成竖排
% \centering %图片局部居中
% \includegraphics[width=0.6\textwidth]{./pictures/BlandAltman.pdf} %此时的图片宽度比例是相对于这个minipage的，不是全局
% \vspace{-3mm}
% \caption{Bland-Altman plot.}
% \label{fig:mymodel_Bland}
% \end{minipage}
% \begin{minipage}[b]{0.59\textwidth} %所有minipage宽度之和要小于1，否则会自动变成竖排
% \centering %图片局部居中
% \includegraphics[width=1\textwidth]{./pictures/mymodel.pdf}%此时的图片宽度比例是相对于这个minipage的，不是全局
% \vspace{-3mm}
% \caption{Curves of predicted HR signals and ground truth HR signals.}
% \label{fig:mymodel}
% \end{minipage}
% \end{figure}

\subsubsection{Memory usage comparison.} \label{sec:4.3.3}
\lsc{This experiment evaluates the memory usage of \sysname compared to baselines.}

\lsc{
\textbf{Setups.}
Using three continuous video clips from UBFC-Phys on Raspberry Pi 4B, with durations of 3 minutes, we monitor the memory usage during the inference process.
For other baselines, as the preprocessing steps couldn't be performed on Raspberry Pi 4B, preprocessed data are used as their input.
We use the rPPG-Toolbox for fair testing.
} \re{The memory usage data of the devices is collected by using the professional tool called dstat.}

\begin{figure}[t]
    \centering
    \includegraphics[width=0.54\textwidth]{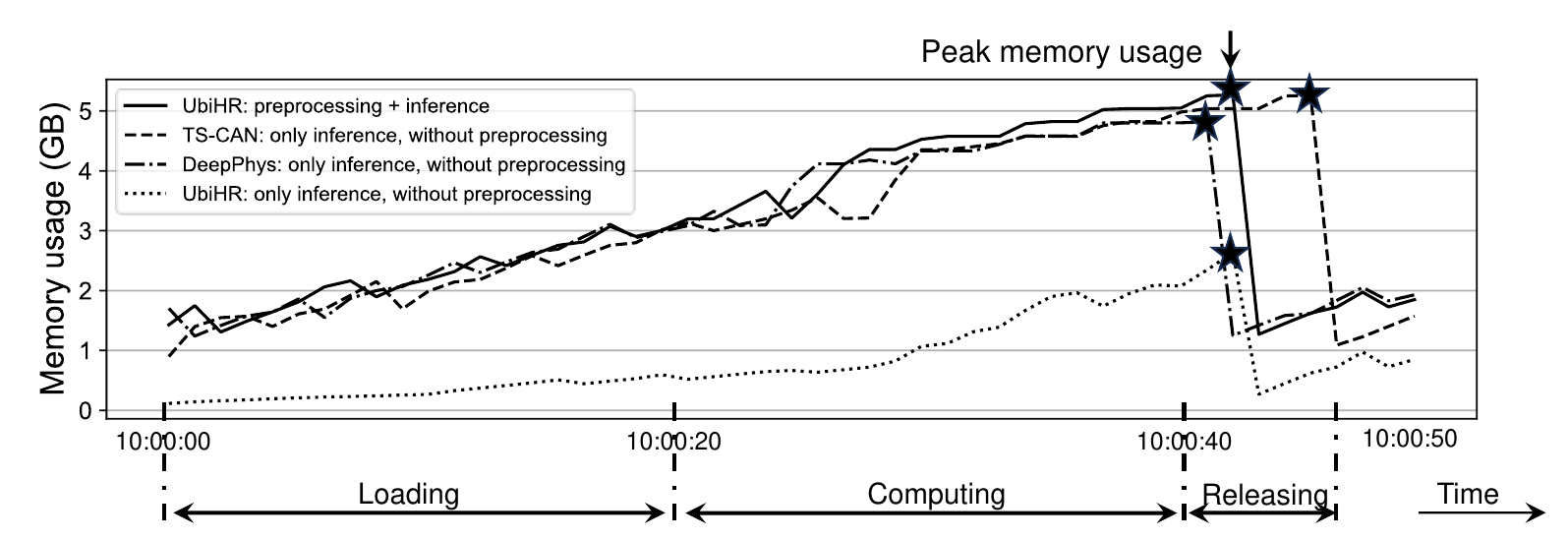}
    \caption{Memory usage comparison of DeepPhys, TS-CAN, and \sysname on Raspberry pi 4B.}
    \vspace{-3mm}
    \label{fig:mem}
\end{figure}

\textbf{Results.}
\lsc{
Figure \ref{fig:mem} illustrates the memory usage of TS-CAN and DeepPhys, which are used solely for inference due to the high-cost preprocessing can not fit into embedded devices. 
While we show the memory usage of both inference and the entire end-to-end process with \sysname.
\textit{First}, UbiHR's memory usage of inference reveals the lowest memory usage.
\textit{Second}, \sysname's end-to-end process exhibits an average memory usage of 2.32GB, which is 1.25\% higher than TS-CAN and 15.71\% higher than DeepPhys.
This indicates that UbiHR's on-device pre-processing and multi-branch spatio-temporal model are low-cost.
\textit{Third}, all three methods exhibit periodic memory usage fluctuations during continuous operation, gradually increasing and decreasing at certain intervals. 
For TS-CAN and DeepPhys, this is due to the need to read pre-processed data in npy format before inference. 
UbiHR conducts on-device pre-processing and inference, requiring reading frames and storing them in memory until a predefined size is reached, at which the memory is released.
}

\subsubsection{Latency comparison}
\lsc{This experiment compares the latency of \sysname and different baseline methods.}

\textbf{Setup.}
\lsc{
The baseline and data setups are consistent with the previous one. 
We test the latency across three devices: Raspberry Pi 4B, Jetson Orin, and Jetson AGX. 
We measure the FPS (Frames Per Second) as the metric.
}

\textbf{Results.} 
\lsc{Table \ref{tab:latency} shows the results. 
\textit{First}, \sysname exhibits the fastest inference speed on all devices. 
On the Raspberry Pi, it shows a 9.06\% improvement over DeepPhys and a 5.24\% improvement over TS-CAN. 
On the Orin and AGX devices, \sysname achieves speed improvements of 60.88\% and 9.55\% respectively over DeepPhys and 7.19\% and 5.32\% over TS-CAN.
\textit{Second}, as device performance increases, \sysname demonstrates the most significant increase in inference speed. From the Raspberry Pi to the AGX, its speed increases by 171.7fps, compared to 156.6fps for DeepPhys and 163fps for TS-CAN. 
This improvement is attributed to the lightweight and parallel spatio-temporal model structure, reducing computing costs.
}

% First, on the same devices, \sysname demonstrated the fastest inference speed. On the Raspberry Pi, we observed a 9.06\% speed improvement compared to DeepPhys and a 5.24\% improvement compared to TS-CAN. On the Orin and AGX devices, \sysname achieved speed improvements of 60.88\% and 9.55\% respectively compared to DeepPhys, and 7.19\% and 5.32\% improvements respectively compared to TS-CAN.
% Second, as device performance improved, \sysname demonstrated the fastest increase in inference speed. From the Raspberry Pi to the AGX, its speed increased by a remarkable 171.7fps, while DeepPhys and TS-CAN increased by 156.6fps and 163fps respectively.
% We attribute this improvement to the efficient spatio-temporal network, which enhances computational efficiency, and the lightweight nature of the parallel network structure, which reduces computational load.

\begin{table}[]
        \centering
        \scriptsize
        \captionsetup[table]{width=0.8\textwidth}
        \captionof{table}{Inference latency comparison of DL-based baselines and \sysname \lsc{on three devices}.} % 使用\captionof命令指定表格标题
        \vspace{-3mm}
        \begin{tabular}{|c|ccc|}
        \hline
        \multirow{2}{*}{\textbf{Method}} & \multicolumn{3}{c|}{\textbf{Frame number inferred per second (FPS) on different device}}                      \\ \cline{2-4} 
                                 & \textbf{Raspberry pi}     & \textbf{Jetson Orin}      & \textbf{Jetson AGX} \\ \hline
        \textbf{DeepPhys~\cite{chen2018deepphys}}                         & \multicolumn{1}{c|}{27.6} & \multicolumn{1}{c|}{61.1} & 184.2               \\ \hline
        \textbf{TS-CAN~\cite{liu2020multi}}                           & \multicolumn{1}{c|}{28.6} & \multicolumn{1}{c|}{91.7} & 191.6               \\ \hline
        \textbf{\sysname}                            & \multicolumn{1}{c|}{30.1} & \multicolumn{1}{c|}{98.3} & 201.8               \\ \hline
        \end{tabular}
        % \vspace{-3mm}
        \vspace{-3mm}
        \label{tab:latency}
\end{table}

\subsection{\lsc{Performance over Diverse Light Conditions}}
\lsc{This experiment assesses the sensing stability of \sysname and three DL-based baseline methods across four real-world scenarios. Varying light conditions introduce different spectra and frequencies, leading to diverse signal-to-noise ratios.  
User motions further lead to noises, challenging the stability of heart rate sensing performance.
}

\begin{figure}[t]
    \centering
    \includegraphics[width=0.55\textwidth]{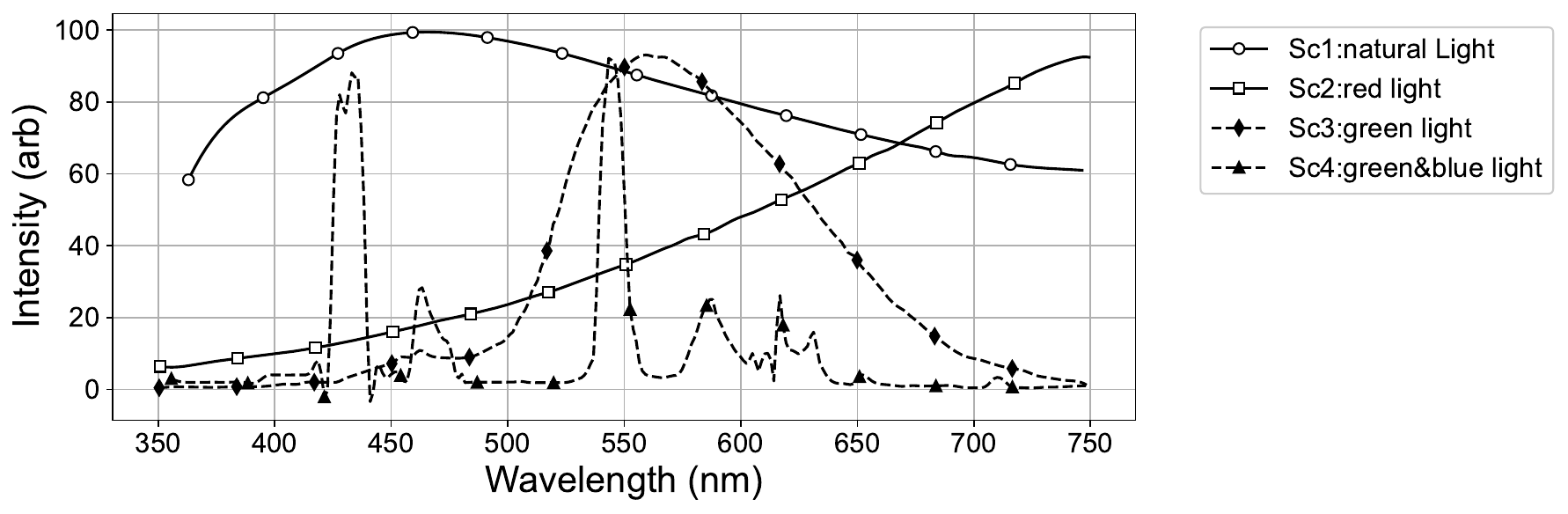}
    \vspace{-3mm}
    \caption{\lsc{Spectrum illusration of different light conditions}.}
    \vspace{-3mm}
    \label{fig:df_light}
\end{figure}

\textbf{Setups.}
\lsc{
We adopt four light conditions that users always encounter, \ie natural light (Sc1), red light (Sc2), green light (Sc3), and mixed blue-green light (Sc4).
These scenarios represent various ubiquitous application environments such as natural outdoor lighting, red light resembling incandescent bulbs, green light resembling fluorescent lamps, and a mix of green and blue light resembling LED lights.
To simulate these conditions, we applied different spectral filters to videos captured under natural lighting (Sc1) using OpenCV.
These videos also feature diverse motions from users.
}

\begin{figure}[t]
\centering  %居中
\subfigure[Sc1: natural light]{   %第一张子图
\begin{minipage}{3.6cm}
\centering    %子图居中
\includegraphics[scale=0.32]{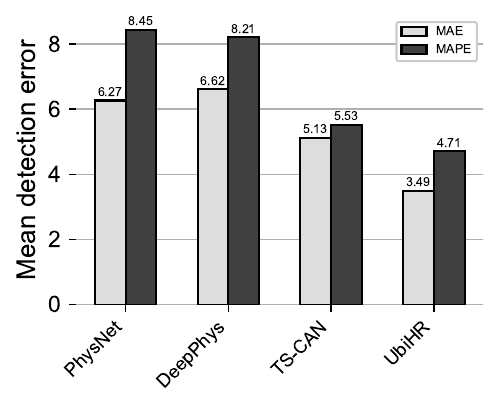}
\end{minipage}
}
\subfigure[Sc2: red light]{ %第二张子图
\begin{minipage}{3.6cm}
\centering    %子图居中
\includegraphics[scale=0.32]{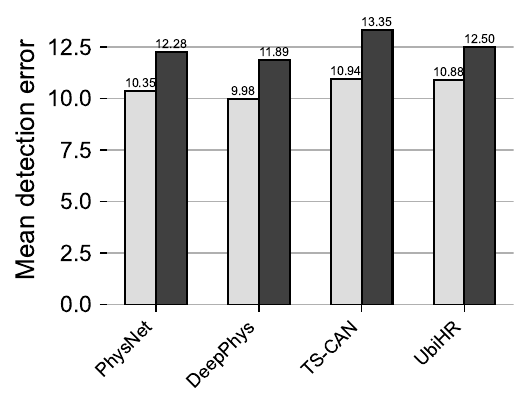}
\end{minipage}
}
\subfigure[Sc3: green light]{ %第三张子图
\begin{minipage}{3.6cm}
\centering    %子图居中
\includegraphics[scale=0.32]{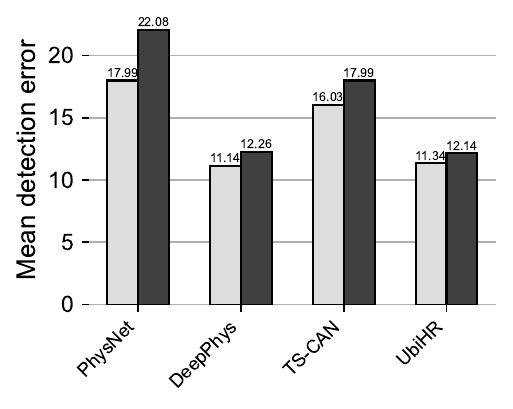}
\end{minipage}
}
\subfigure[Sc4: green\&blue light]{ %第四张子图
\begin{minipage}{3.6cm}
\centering    %子图居中
\includegraphics[scale=0.38]{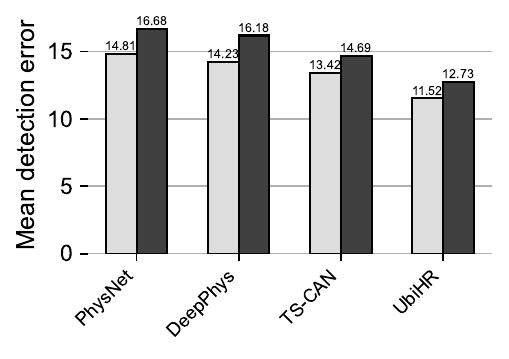}
\end{minipage}
}
\vspace{-3mm}
\caption{Accuracy comparison of three DL-based baselines and \sysname under diverse light conditions.}  
%大图名称
\vspace{-3mm}
\label{fig:df_error}    %图片引用标记
\end{figure}

\begin{figure}[t]
\centering  %居中
\subfigure[Sc1: natural light]{   %第一张子图
\begin{minipage}{3.6cm}
\centering    %子图居中
\includegraphics[scale=0.31]{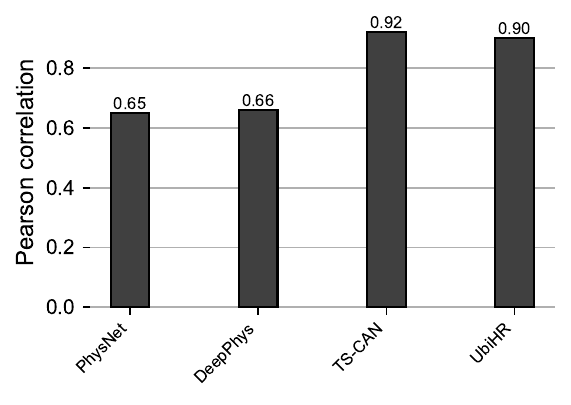}
\end{minipage}
}
\subfigure[Sc1: red light]{ %第二张子图
\begin{minipage}{3.6cm}
\centering    %子图居中
\includegraphics[scale=0.31]{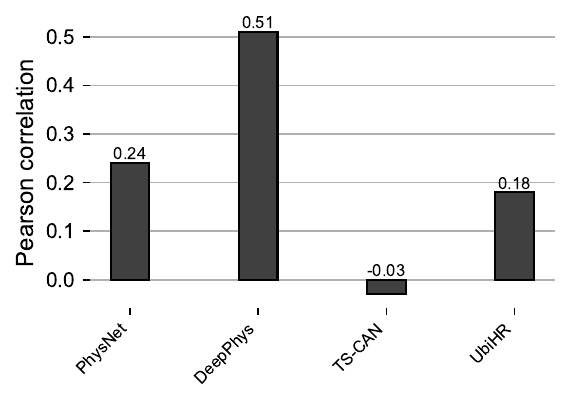}
\end{minipage}
}
\subfigure[Sc1: green light]{ %第三张子图
\begin{minipage}{3.6cm}
\centering    %子图居中
\includegraphics[scale=0.31]{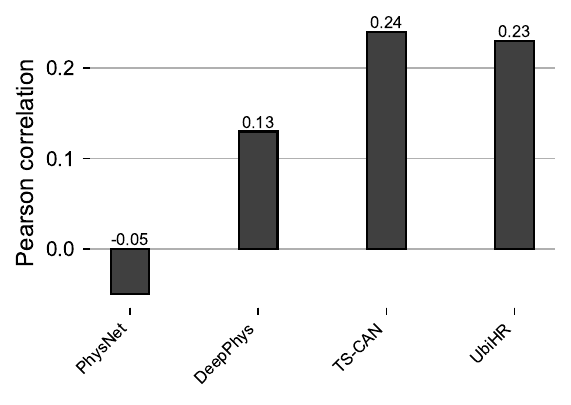}
\end{minipage}
}
\subfigure[Sc1: green\&blue light]{ %第四张子图
\begin{minipage}{3.6cm}
\centering    %子图居中
\includegraphics[scale=0.31]{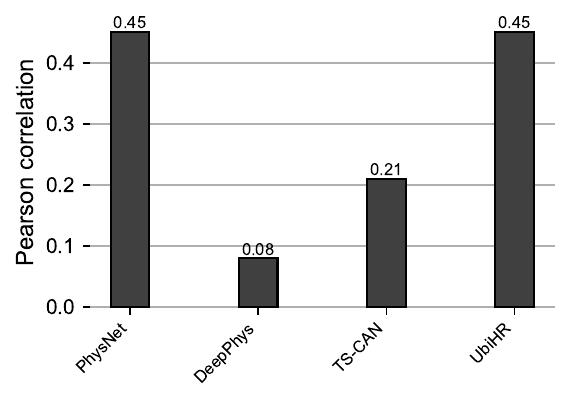}
\end{minipage}
}
\vspace{-3mm}
\caption{Pearson correlation comparison of three baselines and UbiHR under diverse light conditions.}
\vspace{-3mm}
%大图名称
\label{fig:df_pearson}    %图片引用标记
\end{figure}

\begin{figure}[t]
\centering  %居中
\subfigure[Sc1: natural light]{   %第一张子图
\begin{minipage}{3.6cm}
\centering    %子图居中
\includegraphics[scale=0.31]{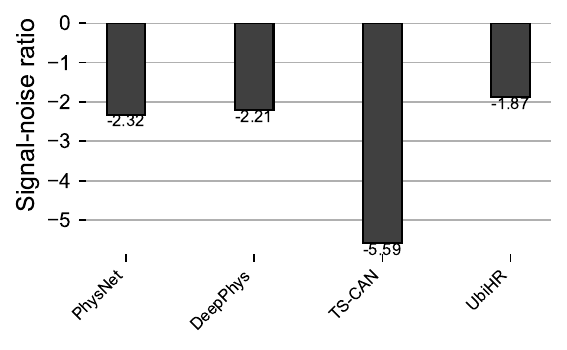}
\end{minipage}
}
\subfigure[Sc1: red light]{ %第二张子图
\begin{minipage}{3.6cm}
\centering    %子图居中
\includegraphics[scale=0.31]{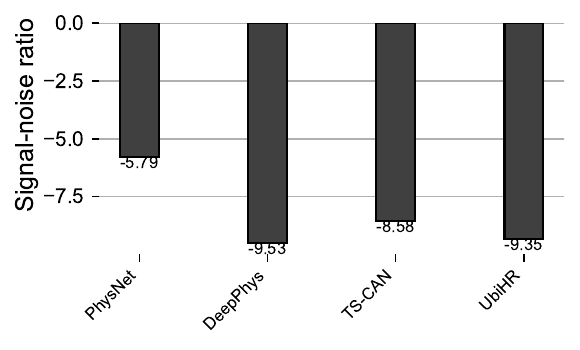}
\end{minipage}
}
\subfigure[Sc1: green light]{ %第三张子图
\begin{minipage}{3.6cm}
\centering    %子图居中
\includegraphics[scale=0.31]{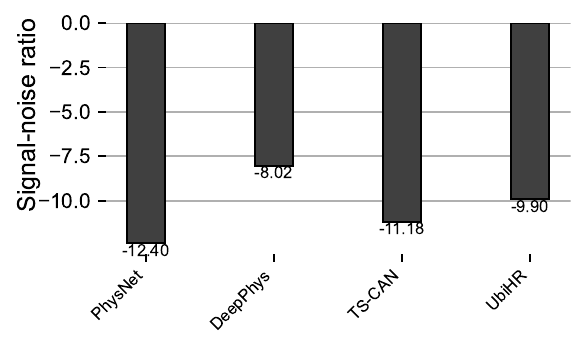}
\end{minipage}
}
\subfigure[Sc4: green\&blue light]{ %第四张子图
\begin{minipage}{3.6cm}
\centering    %子图居中
\includegraphics[scale=0.31]{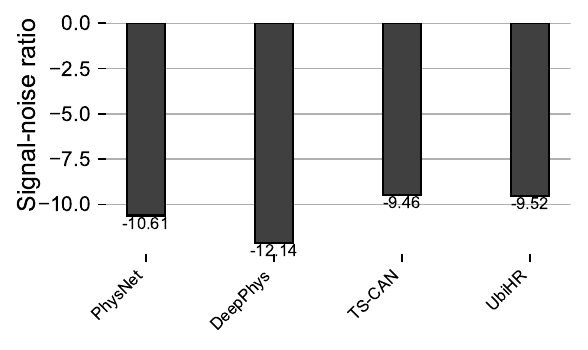}
\end{minipage}
}
\vspace{-3mm}
\caption{Signal-noise ratio comparison of three baselines and UbiHR under diverse light conditions.}  
\vspace{-3mm}
%大图名称
\label{fig:df_snr}    %图片引用标记
\end{figure}

\textbf{Results.} 
\lsc{
As shown in \figref{fig:df_light}, natural light offers a broad spectrum with uniform frequencies, resulting in ideal optical reflection from the skin. 
In contrast, artificial indoor light sources emit specific frequencies.
\textit{First}, in \figref{fig:df_error}, UbiHR shows the best stability with a minimal MAE fluctuation of 0.64, outperforming DeepPhys (MAE fluctuation: 4.25), which excels under red and green lights but struggles under blue-green. 
PhysNet and TS-CAN exhibit higher fluctuations (7.74 and 5.09, respectively). 
UbiHR's stability is attributed to its long-range multi-branch network, adept at capturing heart rate variations and reducing noise impact even with drastic lighting changes. 
\re{Also, all methods perform better } under red light but face challenges under green due to variations in blood's light absorption properties, with red light experiencing less interference and green light fluctuations causing more significant optical changes, affecting accuracy.}
\lsc{\textit{Second}, in terms of the correlation between predicted and actual heart rate signals, UbiHR demonstrates superior stability. 
PhysNet and TS-CAN exhibit negative correlations under green and red light conditions, respectively.
DeepPhys shows a weak correlation coefficient of only 0.08 under blue-green mixed light, also considered unacceptable. 
While UbiHR not only shows better correlations across conditions but also maintains relatively small fluctuations, affirming its adeptness at learning intrinsic heart rate signal variations.
Third, we assess the signal-to-noise ratio (SNR), indicating a method's noise resistance, where lower values imply stronger noise resistance, considering the weak BVP signal against strong noise. UbiHR excels in SNR performance under natural light, outpacing TS-CAN by 67.6\%. 
Also, it maintains stability under other lighting conditions, avoiding sudden/significant performance drops.
}
% However, under other lighting conditions, all models experience a notable SNR decline. 
% Interestingly, there's a correlation between higher SNR and higher error rates overall, reinforcing our initial findings.

\subsection{Performance with Diverse Input Range} \label{sec:4.5}
\lsc{
This experiment confirms the effectiveness of \sysname's long-range spatio-temporal model compared to baseline methods relying on short-range adjacent frames. 
As mentioned in \bhy{\secref{sec:3.4}}, previous methods always utilize a single-branch structure, such as TS-CAN~\bhy{\cite{liu2020multi}}, restricting their capability to establish mappings from videos to heart rate signals only within short ranges. 
While \sysname adopts a multi-branch parallel spatio-temporal model, enabling simultaneous capture of spatio-temporal relationships across short and long ranges.
}

% Only accuracy and robustness tests in different environments still cannot fully demonstrate the inference capability of \sysname in the long-range. 
% This experiment tests the performance of modeling long-range frame sequences, which reflects the robustness of the model to varying input lengths. 

%As disscussed in Sec 3,先前的方法通常采用单一分支的结构，在相邻帧之间进行时空建模，这使得它们只能建立short-range上光线变化与心率信号的映射,例如TS-CAN就采用了这种结构，而UbiHR使用的多分支并行时空建模网络可以兼顾short-range和long-range的时空建模

\textbf{Setup.}
%我们在Device $D_1$上使用同一段输入视频进行测试。我们将每次输入TS-CAN和UbiHR的帧序列数量从5帧开始递增直到30帧，测试不同输入帧数量对模型推理准确度的影响。
\lsc{
We perform experiments on Device $D_1$ using the same three input video segments as discussed in \bhy{\secref{sec:4.3.3}}. 
We vary the number of frame sequences inputted to TS-CAN and \sysname, ranging from adjacent 5 to 30 frames, to investigate the influence of different input quantities on inference accuracy. 
Each input undergoes 5 times to show the average value.
}

\begin{figure}[t]
\centering %图片全局居中
%并排几个图，就要写几个minipage
\begin{minipage}[b]{0.32\textwidth} %所有minipage宽度之和要小于1，否则会自动变成竖排
\centering %图片局部居中
\includegraphics[width=0.7\textwidth]{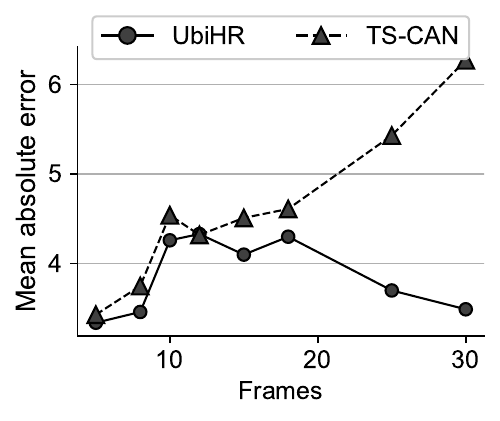} %此时的图片宽度比例是相对于这个minipage的，不是全局
\vspace{-2mm}
\caption{Impact of input frame number.}
\vspace{-2mm}
\captionsetup{width=0.75\textwidth}
\label{fig:long_term}
\end{minipage}
\begin{minipage}[b]{0.32\textwidth} %所有minipage宽度之和要小于1，否则会自动变成竖排
\centering %图片局部居中
\includegraphics[width=0.7\textwidth]{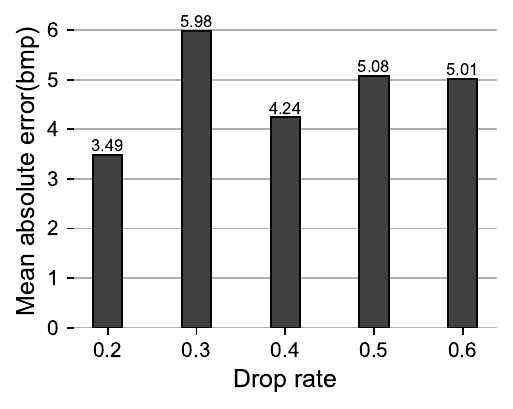}%此时的图片宽度比例是相对于这个minipage的，不是全局
\captionsetup{width=0.75\textwidth}
\vspace{-2mm}
\caption{Impact of drop rates.}
\vspace{-2mm}
\label{fig:droprate}
\end{minipage}
\begin{minipage}[b]{0.33\textwidth} %所有minipage宽度之和要小于1，否则会自动变成竖排
\centering %图片局部居中
\includegraphics[width=0.7\textwidth]{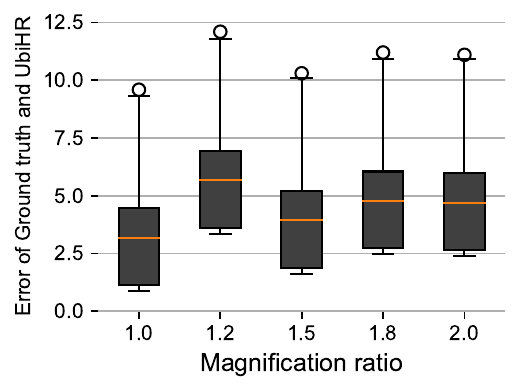} %此时的图片宽度比例是相对于这个minipage的，不是全局
\vspace{-2mm}
\caption{Impact of face box size.}
\vspace{-2mm}
\captionsetup{width=0.75\textwidth}
\label{fig:facebox}
\end{minipage}
\end{figure}

\textbf{Results.} 
\lsc{\figref{fig:long_term} illustrates the changes in heart rate prediction error for \sysname and TS-CAN with increasing input frame ranges. 
TS-CAN exhibits a gradual increase in error as the input frame range rises, whereas \sysname maintains a relatively stable error rate between 3.5\% and 5\%. 
The difference in error is minimal when the input frame range is around 12 frames for \bhy{TS-CAN} but gradually widens thereafter. 
Notably, when utilizing a range of 30 frames as input, \sysname achieves an error reduction of approximately 44\% compared to TS-CAN.
Therefore, in TS-CAN~\cite{liu2020multi}, the default input is limited to 10 frames.
}

\begin{table}[t]
\scriptsize
\caption{Performance of with/without soft-attention mask in \sysname.}
\vspace{-3mm}
\begin{tabular}{|c|cc}
\hline
\multirow{2}{*}{\textbf{Error metrics}} & \multicolumn{2}{c|}{\textbf{Abulation set}}                                             \\ \cline{2-3} 
                                  & \multicolumn{1}{c|}{With attention mask} & \multicolumn{1}{c|}{Without attention mask} \\ \hline
MAE                               & \multicolumn{1}{c|}{3.49}                & \multicolumn{1}{c|}{6.81}                  \\ \hline
MAPE                              & \multicolumn{1}{c|}{4.65}                & \multicolumn{1}{c|}{8.67}                  \\ \hline
RMSE                              & \multicolumn{1}{c|}{6.01}                & \multicolumn{1}{c|}{13.62}                  \\ \hline
Pearson                           & \multicolumn{1}{c|}{0.90}                & \multicolumn{1}{c|}{0.45}                   \\ \hline
SNR                               & \multicolumn{1}{c|}{-2.21}               & \multicolumn{1}{c|}{-4.61}                 \\ \hline
\end{tabular}
\vspace{-2mm}
\label{tab:mask}
\end{table}

\subsection{Performance on Diverse Testing Data}
This experiment validates the inference speed and robustness of \sysname in diverse testing data, on the smartphone  ($D_4$). 
We deploy the model pre-trained on UBFC-rppg, and test it on three different datasets (UBFC-Phys, SCAMPS, and UCLA-rPPG). 
To simulate real-world scenarios, we maintain the resource-competing processes on the mobile phone. 
% Although the iQOO9 device is equipped with a mobile GPU, we primarily conduct tests on the CPU for the sake of portability and scalability. 
The results summarized in \tabref{tab:iqoo9} indicate that \sysname achieves an inference speed of 12.4ms per frame on the Snapdragon 8 Gen1 CPU, meeting real-time detection requirements in practical scenarios. 
Moreover, in terms of inference accuracy, \sysname achieves MAE values of 4.13, 5.07, and 5.28 on the three different datasets, respectively, demonstrating robustness in real-world application scenarios.

\begin{table}[t]
\scriptsize
\caption{Performance on mobile phone iQOO9 of \sysname.}
\vspace{-3mm}
\begin{tabular}{|c|c|ccc|}
\hline
\multirow{2}{*}{\textbf{Device}} & \multirow{2}{*}{\textbf{Latency}} & \multicolumn{3}{c|}{\textbf{Testing error(MAE)}}                        \\ \cline{3-5} 
                                 &                                   & \multicolumn{1}{c|}{UBFC-Phys} & \multicolumn{1}{c|}{SCAMPS} & UCLA-rPPG \\ \hline
iQOO 9                           & 12.4ms                            & \multicolumn{1}{c|}{4.13}      & \multicolumn{1}{c|}{5.07}   & 5.28      \\ \hline
\end{tabular}
\label{tab:iqoo9}
\end{table}

\subsection{Micro-benchmark and Ablation studies}

\subsubsection{Impact of Drop Rate}
\lsc{We examine the influence of various Drop Rates during \sysname's model training. 
As mentioned in \secref{sec:2DCNN}, we utilize 2D convolution kernels and the drop rate is an important parameter that affects its learning effectiveness.
As depicted in \figref{fig:droprate}, the optimal Drop Rate in \sysname's training process is 0.2. 
Further elevating this parameter diminishes both model performance and robustness.
}

\subsubsection{Impact of Face Box Size}
\lsc{We test the influence of the size of the face bounding box formed by key point detection on \sysname's accuracy. 
As depicted in \figref{fig:facebox}, continuously increasing the enlargement ratio of the face bounding box results in decreased performance and stability of \sysname. 
This is because enlarging the face bounding box range introduces background noise, thus affecting accuracy.
}

\subsubsection{Impact of Soft-attention Mask}
\lsc{As discussed in \secref{sec:attention}, directly integrating the spatio-temporal module to 2D convolution introduces extra spatio-temporal noise, diverting the network's attention to pixels devoid of physiological signals. 
Thus, we introduce the soft-attention mask as a correction. 
Table \ref{tab:mask} compares the performance of \sysname on the UBFC-Phys test data with and without the soft-attention mask. 
It's evident that without the soft-attention mask, \sysname's accuracy notably declines across all 5 metrics.
}

\subsection{Case study}

\begin{figure}[t]
    \centering
    \begin{minipage}{0.6\linewidth}
        \centering
        \subfigure[$S_1$: dormitory]{
            \includegraphics[scale=0.02]{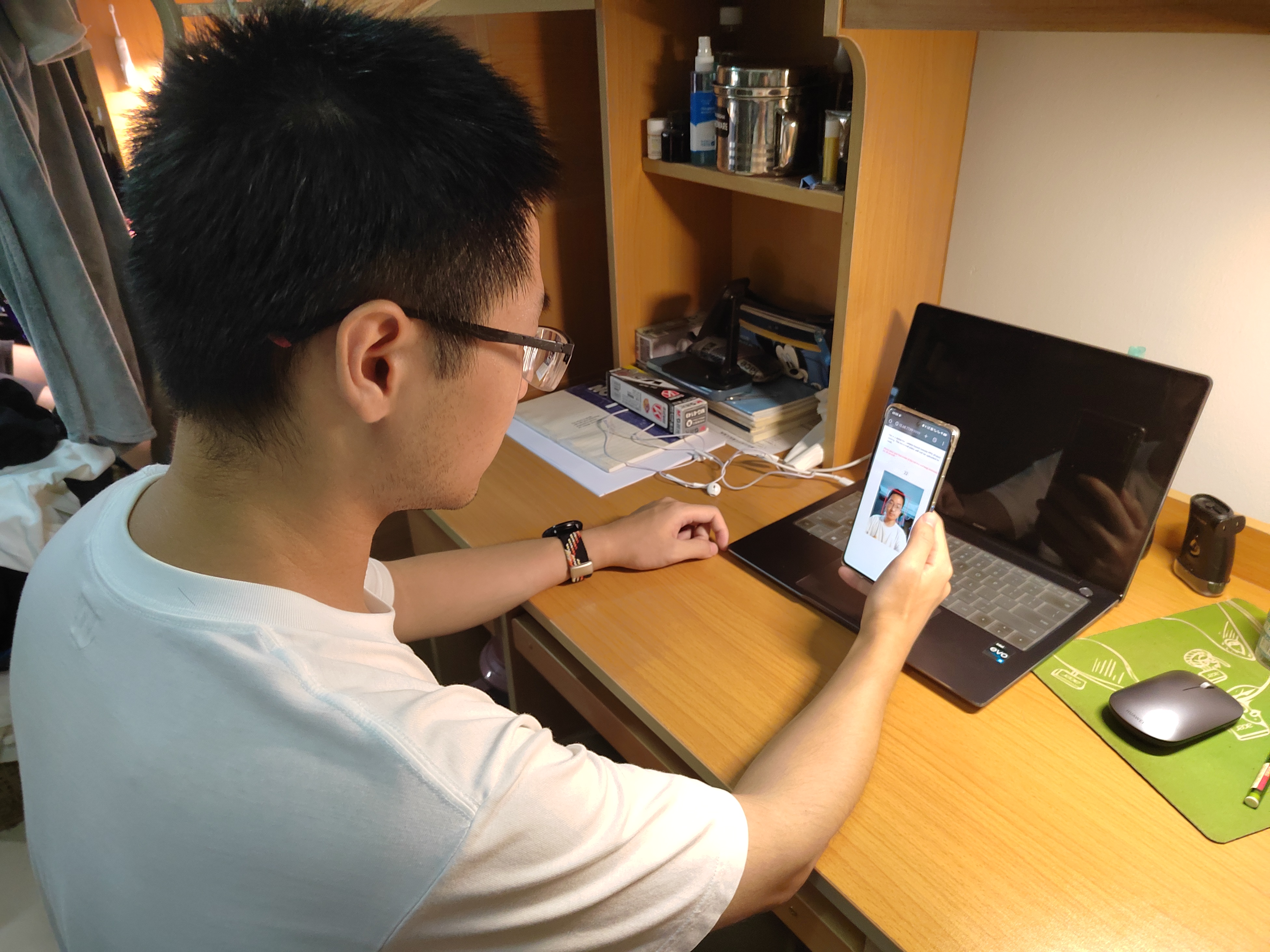}
        }
        \subfigure[$S_2$: meeting room]{
            \includegraphics[scale=0.046]{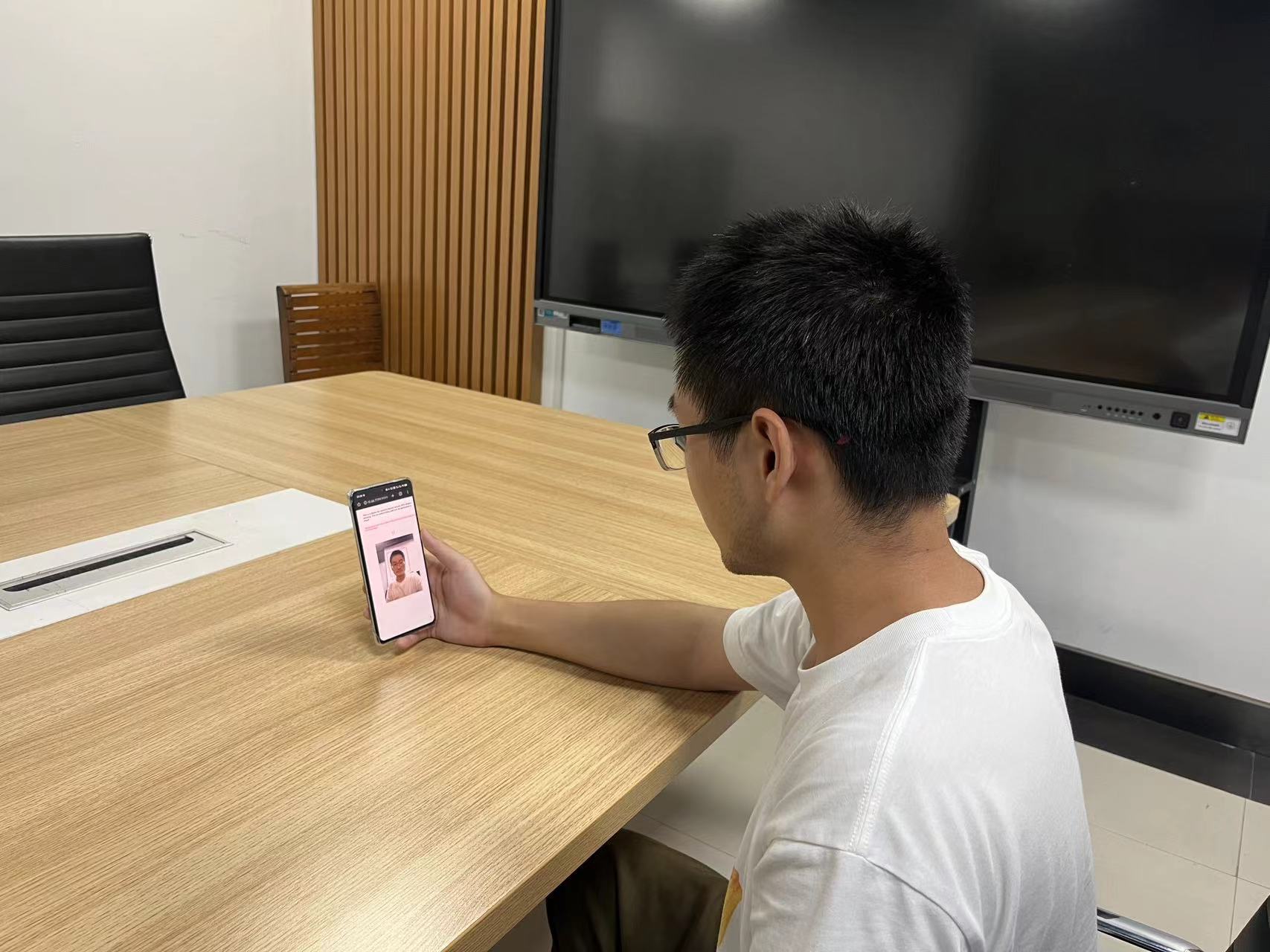}
        }
        \subfigure[$S_3$: cafe]{
            \includegraphics[scale=0.02]{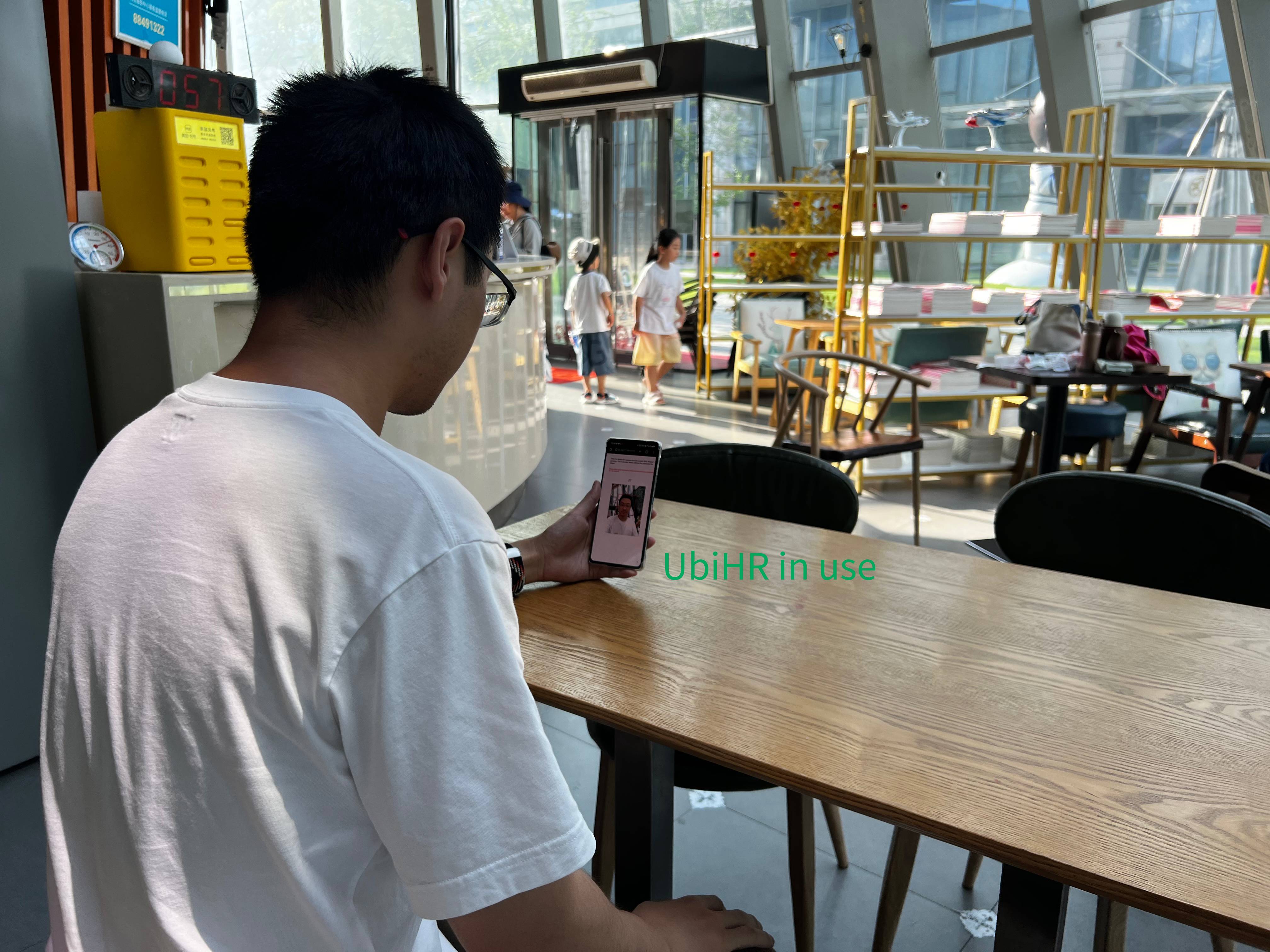}
        }
        \vspace{-3mm}
        \caption{\re{Three real-world case studies in real-world scenarios.}}
        \vspace{-3mm}
        \label{fig:scenarios}
    \end{minipage}
    \hfill
    \begin{minipage}{0.39\linewidth}
        \centering
        \includegraphics[width=0.6\linewidth]{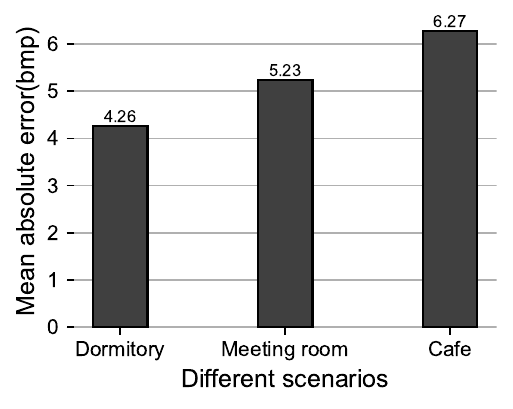}
        \vspace{-3mm}
        \caption{\re{Performance in different scenarios}}
        \vspace{-3mm}
        \label{fig:result}
    \end{minipage}
\end{figure}

\re{To investigate the performance of \sysname in various real-world scenarios, we conduct tests using a smartphone ($D_4$) across multiple real-world settings. 
As shown in \TODO{\figref{fig:scenarios}}, we select three scenarios for testing:
\textit{i)} Dormitory: features stable fluorescent lighting with minimal background pedestrian traffic, making it close to an ideal experimental environment.
\textit{ii)} Meeting room: similar to the dormitory, it has stable lighting from natural light or LED lamps but is larger and more complex, with potential interference from objects and moving people.
\textit{iii)} Cafe: the most complex environment, combining natural, fluorescent, and LED lighting, with a physically intricate setting and high pedestrian traffic.
We assess \sysname's inference accuracy in these progressively complex real-world scenarios.
We use the front camera of the smartphone to record a 10-minute video, and then performed heart rate monitoring through the \sysname system deployed on the smartphone while the user wearing professional heart rate monitoring equipment. 
The results are shown in the \figref{fig:result}, in different scenarios, the MAE of \sysname are 4.26, 5.23, and 6.27 respectively. 
This is consistent with our expectation, that is, the more complex the lighting conditions and the environment, the greater the noise in \sysname's inference, and the higher the error rate. 
Even so, \sysname still maintaines a relatively high accuracy in the most complex scenario.
}

\begin{figure}[t]
\centering 
\includegraphics[width=0.8 \textwidth]{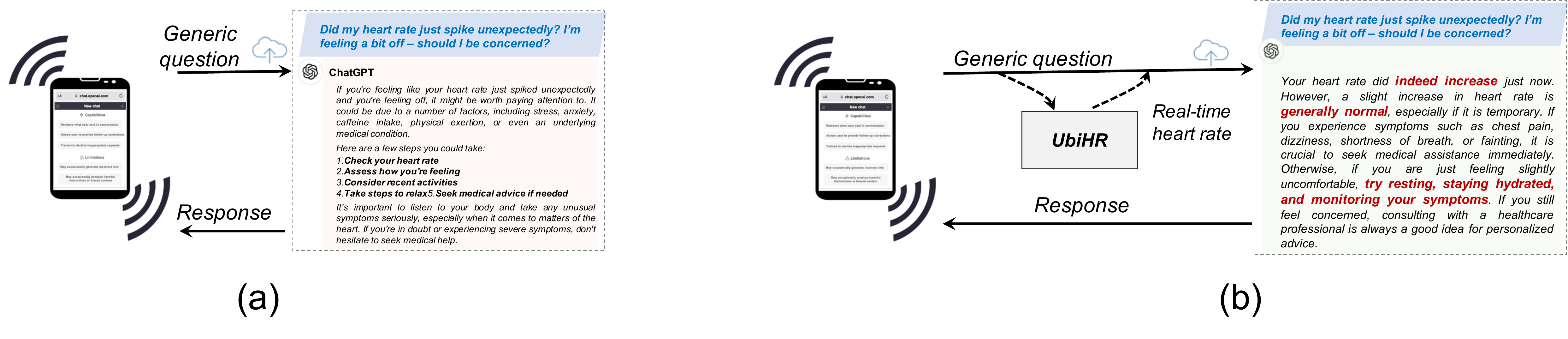}
\vspace{-3mm}
\caption{Comparison of ChatGPT's advice to heart health issues: (a) Generic healthcare prompt; (b) Incorporating \sysname's real-time heart rate sensing result into prompt}
\vspace{-3mm}
\label{fig:case_study}
\end{figure}

Large Language Models (LLMs), exemplified by OpenAI's ChatGPT, have demonstrated success in the healthcare field~\cite{singhal2023large, xiong2023doctorglm}. 
However, when addressing cardiovascular diseases, LLMs often provide vague answers due to the lack of accurate and real-time user physiological data.
To showcase the application value and performance of \sysname, we conduct a practical case study using ChatGPT on mobile devices. 
We compare the outputs of ChatGPT under two scenarios: direct user inquiries for health advice and integration of real-time heart rate sensing results from \sysname into the query prompt. 
% The comparison, illustrated in Fig. \ref{fig:case_study}, highlights the differences in outcomes.
In Fig. \ref{fig:case_study}a, when a user seeks healthcare advice from ChatGPT without real-time physiological data, the response lacks specificity. 
In Fig. \ref{fig:case_study}b, with the integration of real-time heart rate data from \sysname, the specificity and assistance of the response are significantly improved.

% Large Language Models (LLMs), represented by OpenAI's ChatGPT, have demonstrated success in healthcare field~\cite{singhal2023large，xiong2023doctorglm}.
% However, when it comes to cardiovascular diseases, LLMs often provide vague answers due to the lack of accurate and realtime user physiological data.

% 因此，我们做了一个实际的case study, 采用手机端chatgpt问答, 对比用户直接询问健康建议，和将\sysname实时hear rate sensing results 集成至问答prompt两种情况下的chatgpt输出，以体现\sysname的应用价值和性能表现。
% \bhy{The comparison presented in Fig. \ref{fig:case_study} contrasts the results. 
% When a user feels uncomfortable and consults ChatGPT (\figref{fig:case_study}a), the response lacks specificity, indicating that the LLM provides a generic reply without promptly identifying the root cause of the health concern to offer appropriate advice.
% %
% Conversely, in \figref{fig:case_study}b, with the addition of real-time heart rate data from \sysname, the 专用性和帮助 of the answer is obviously improved. 
% Specifically, the LLM can summarize the patterns of the user's heart rate changes based on their realtime heart rate sensing results, and then determine whether the situation is abnormal or normal. 
% This demonstrates the 价值 of \sysname in real-time monitoring of users' heart rates via ubiquitous devices.

\begin{figure}[t]
\centering %图片全局居中
%并排几个图，就要写几个minipage
\vspace{-3mm}
\begin{minipage}[b]{0.4\textwidth} %所有minipage宽度之和要小于1，否则会自动变成竖排
\centering %图片局部居中
\includegraphics[width=0.59\textwidth]{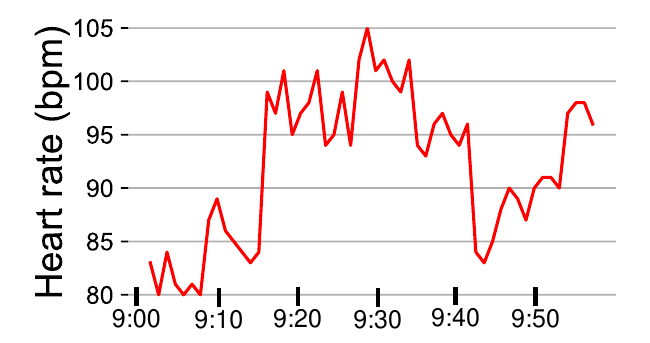} %此时的图片宽度比例是相对于这个minipage的，不是全局
\caption{Real-time heart rate of the participants.}
\vspace{-4mm}
\captionsetup{width=0.6\textwidth}
\label{fig:hr}
\end{minipage}
\begin{minipage}[b]{0.5\textwidth} %所有minipage宽度之和要小于1，否则会自动变成竖排
\centering %图片局部居中
\includegraphics[width=0.5\textwidth]{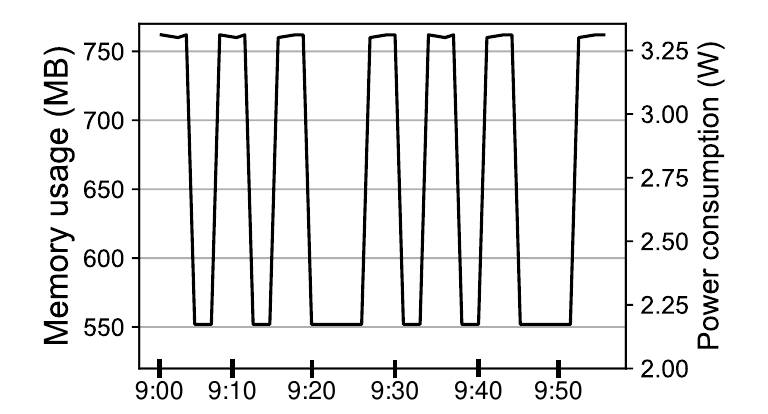}%此时的图片宽度比例是相对于这个minipage的，不是全局
\captionsetup{width=1\textwidth}
\caption{Changes in memory usage and power consumption.}
\vspace{-4mm}
\label{fig:mobile_mem}
\end{minipage}
\end{figure}

We conduct tests to evaluate the system overhead of deploying \sysname on a smartphone ($D_4$). 
Using the front-facing camera, we perform a 55-minute long-duration test to monitor participants' heart rate fluctuations, as illustrated in Fig. \ref{fig:hr}.
In terms of memory usage, we observe an average usage of 660.5MB, with a peak usage of 762MB. 
The front-facing camera accounts for approximately 532MB, while \sysname occupies around 230MB.
Regarding power consumption, the \sysname system exhibits an average power consumption of 2.77W, with a peak consumption of 3.3W. 
The front-facing camera consumes approximately 2.2W, while \sysname's inference consumes around 1.1W. 
Thanks to the adaptive duty cycling mechanism, \sysname reduces memory usage and power consumption by 13.3\% and 19\% respectively compared to continuous inference. 
This optimization enables \sysname to operate with low resource usage and power cost, making it suitable for deployment on various mobile platforms for long-term heart rate sensing.

% Specifically, we conduct tests on the system overhead of deploying \sysname on a smartphone ($D_4$). 
% We perform an 55min test using the front-facing camera on Device $D_4$. 
% The fluctuation state of participants' heart rate is illustrated in \figref{fig:hr}.
% First, in terms of memory usage, the average usage is 660.5MB, with a peak usage of 762MB. The front-facing camera accounts for approximately 532MB, while \sysname occupies around 230MB.
% Second, regarding power consumption, the \sysname system has an average power consumption of 2.77W and a peak power consumption of 3.3W. The front-facing camera consumes approximately 2.2W, while \sysname's inference consumes around 1.1W. Thanks to the duty cycling mechanism, \sysname reduces memory usage and power consumption by 13.3\% and 19\% respectively compared to continuous inference. This enables \sysname to operate with low resource utilization and low power consumption, making it suitable for deployment on various mobile platforms.

%% file: body/6-related-works.tex
\section{RELATED WORK}\label{sec:related}
Our work is related to the following categories of research.

\re{\subsection{Mobile Health Monitoring Powered by Machine Learning}
Biometric signal sensors powered by machine learning, focusing on important biosignals: electrocardiogram (ECG), electromyogram (EMG), electroencephalogram (EEG), and photoplethysmogram (PPG) has already achieved widespread research and application.~\cite{kim2024recent}. Furthermore, the diagnostic and monitoring capabilities for diseases such as Inflammatory Bowel disease(IBD)~\cite{hirten2024longitudinal} and Breast diseases~\cite{abrantes2023external} have also seen tremendous improvements with the help of deep learning models. Apart from the development of deep learning combined with personalized healthcare research~\cite{calisto2022modeling}, this can be attributed to the development of various wearable or non-invasive health monitoring and assistive devices~\cite{kazanskiy2024review}. The widespread adoption and development of cameras have also provided the hardware foundation for researching video-based non-contact heart rate monitoring systems.}

\subsection{Heart rate sensing tools}
Contact-based heart rate sensing tools are widely used for heart rate measurement across various crowds to continuously monitor health conditions \cite{kawasaki2021estimating, fedorin2021heart}. 
However, prolonged wearing of contact-based monitors such as finger clips or chest straps could cause discomfort for users and may impact their daily lives. Wearable devices, \eg wristbands \cite{cao2022guard}, eyeglasses \cite{chwalek2023captivates}, and AR/VR headsets \cite{zhang2022personalized, kim2023physiological}, are gradually becoming integrated into people's daily lives. 
For example, EmoTracer \cite{wang2022emotracer} integrates a blood oxygen module and a heart rate module into a wristband-type sensor for acquiring physiological signals. 
\re{
Some studies use smartphone cameras~\cite{siddiqui2016pulse} or other sensors, \eg GPS~\cite{homdee2019enabling}, to perform contact-based measurement or estimation. }
% For example, Siddiqui et al.~\cite{siddiqui2016pulse} proposed an algorithm for heart rate monitoring using only the smartphone camera as the sensor. And Homdee, N. et al.~\cite{homdee2019enabling}propose the use of sensor data from the smartphone, such as  accelerometer data and GPS data to estimate the missing heart rate values.
 Nevertheless, the positioning of devices on different body locations can also lead to variations in measurement results. Therefore, adopting non-contact methods for heart rate sensing will be more convenient and efficient for users.

\subsection{Function-based visual heart rate sensing}
To address the limitations of contact-based heart rate sensing, remote heart rate sensing has garnered increasing attention. Among these approaches, remote photoplethysmography (rPPG) has emerged as an important alternative to existing sensing tools, as it solely relies on a camera \cite{sun2022estimating}. Function-based visual heart rate sensing approaches endeavor to extract heart rate signals from facial videos based on mathematics and designed models, such as independent component analysis (ICA) \cite{wei2017non}, blind source separation (BSS) \cite{poh2010non}, and chrominance model \cite{lewandowskameasuring}. Guo et al. \cite{guo2014physiological} utilize joint BSS to analyze color signal changes from multiple facial sub-regions for remote heart rate sensing. POS \cite{wang2016algorithmic} incorporates the relevant optical and physiological properties of skin reflections, using a projection plane orthogonal to the skin-tone for pulse extraction. Nevertheless, function-based methods lack robustness when detecting in complex surroundings or under obvious changes in lighting conditions. UbiHR enhances the anti-interference capability against real-world sensing scenario noise through a series of heart rate sampling and preprocessing steps, aiming to provide more accurate heart rate sensing results.

\subsection{DL-based heart rate sensing}
Constantly evolving deep learning (DL) techniques have laid the foundation for more accurate and robust visual heart rate sensing \cite{vspetlik2018visual,dark2022heart}. Specifically, CNN-based approaches utilize feature extraction networks to obtain high-quality rPPG signals and then estimate heart rates based on signal features, mainly including 2D CNN \cite{liu2020multi, chen2018deepphys, niu2018synrhythm} and 3D CNN \cite{yu2019remote, perepelkina2020hearttrack}. To capture richer temporal information of rPPG, RNN have been incorporated into heart rate sensing to mitigate sensing data noise \cite{botina2020long}, model the correlation between adjacent frames \cite{niu2019rhythmnet}, and extract global temporal information from continuous facial video frames \cite{huang2021novel}. For example, Lee et al. \cite{lee2020meta} integrate meta learning with LSTM to enhance the adaptability of heart rate sensing models to random variations in facial videos. Furthermore, GAN have been employed to compute higher-quality ROI for heart rate sensing \cite{lu2021dual}. Pulsegan \cite{song2021pulsegan} derives an initial rPPG signal from the delineated ROI, which is subsequently utilized as input to generate realistic rPPG signals. Additionally, the performance of Transformers \cite{yu2022physformer, gupta2023radiant} has been validated in modeling long-term dependencies of rPPG signals. However, these methods necessitate substantial computational and storage resources during both training and inference, posing challenges for their deployment on resource-constrained smart devices. UbiHR builds upon prior research efforts to develop a lightweight heart rate sensing system capable of operating on ubiquitous devices.

\subsection{Resource-efficient on-device heart rate sensing}
Deploying advanced AI models on resource-constrained smart devices for continuous heart rate sensing remains a challenging research endeavor \cite{liu2020multi, liu2022mobilephys}. To make sensing models more lightweight, Comas et al. \cite{comas2022efficient} incrementally combining multiple convolutional derivatives to model rPPG dynamics. RtrPPG \cite{botina2022rtrppg} optimizes inference time through methods such as input size reduction, and RGB to YUB color space transformation. Similarly, Efficientphys \cite{liu2023efficientphys} strives to minimize the computational overhead and inference time of the entire model by minimizing the need for face segmentation, normalization, color space transformation, or any other unnecessary preprocessing steps. Moreover, Liu et al. \cite{liu2024lightweight} combine empirical mode decomposition with channel-wise lightweight CNN to infer heart rate. Some researchers have also utilized the shallow encoder-decoder framework \cite{chowdhury2024lgi} to achieve resource-efficient heart rate sensing. UbiHR follows this research trend of cultivating resource-efficient on-device heart rate sensing systems. It performs real-time heart rate sensing by applying energy-efficient data sampling and concise facial video frame pre-processing techniques.

\subsection{Spatio-temporal DL model}
Spatio-temporal modeling is an essential part of heart rate sensing, as the rPPG signal reflects both subtle local color changes indicating blood volume variations and periodic changes corresponding to the cardiac cycle. Effective spatio-temporal modeling is necessary to identify relevant facial regions and track rPPG evolution. Spatio-temporal DL models primarily leverage combinations of 2D CNN and RNN \cite{lee2020meta}, two-stage CNN \cite{wang2019vision}, and 3D CNN \cite{yu2019remote} (or combined with Temporal Shift Models \cite{lin2019tsm, liu2020multi}) to accurately capture the rPPG signal. For instance, CVD \cite{niu2020video} transforms facial videos into multi-scale spatio-temporal maps, preserving the temporal characteristics of periodic physiological signals while suppressing redundant data. Yu et al. \cite{yu2020autohr} utilize Temporal Difference Convolution (TDC) to capture intrinsic rPPG-aware clues between facial video frames and then use spatio-temporal data augmentation strategies for rPPG extraction. To reduce redundant spatial information, SAM \cite{hu2021robust} transforms long-range spatio-temporal feature maps into short-segment spatio-temporal feature maps, followed by spatio-temporal strip pooling and attention mechanisms to extract rPPG signals. However, since the rPPG signal may span multiple cardiac cycles, long-range spatio-temporal modeling can better capture the long-term dependencies and periodicity in the rPPG signal compared to short-range spatio-temporal modeling, effectively handling heart rate variations and irregular rhythms. To this end, UbiHR designs a simple yet effective preprocessing method based on image differencing, along with a long-range spatio-temporal modeling network, enhancing the robustness of measurements in different environments while ensuring minimal latency.

%% file: body/Conclusion.tex
\section{Conclusion}\label{sec:conclusion}
This paper presents \sysname, a real-time on-device heart rate sensing system on resource-constrained ubiquitous
devices.
\sysname is the first work that realizes end-to-end real-time noise-robust long-range spatio-temporal heart rate sampling, pre-processing, and recognizing.
It includes adaptive duty-cycling facial video sampling, dynamic noise-aware facial image pre-processing, and fast long-range spatio-temporal heart rate recognition.
Evaluations on \bhy{4} datasets and real-world scenarios over four mobile and embedded devices show that \sysname achieves latency reductions of up to 51.2\%, and accuracy improvements of up to 74.2\%.
In the future, we plan to integrate \sysname with the large language model-based
interaction robotics, creating the real-time on-device physiological sensing and interaction system loop.

%% file: body/appendix.tex
\appendix

\section{Appendix}

\subsection{Optical Principles}
In \secref{subsec:preprocess}, we provide the optical principles behind the differential frames, and here we give the detailed proof. The RGB values captured by the cameras is given by:
\begin{equation}
    C_k(t) = I(t) \cdot (v_s(t) + v_d(t)) + v_n(t)
\end{equation}
where $I_(t)$ is the luminance intensity level, modulated by the specular reflection $v_s(t)$ and the diffuse reflection $v_d(t)$. The quantization noise of the camera sensor is captured by camera sensor is captured by $v_n(t)$. We can decompose $v_s(t)$ and $v_d(t)$ into stationary and time-dependent parts:
\begin{equation}
    v_d(t) = u_d \cdot d_0 + u_p \cdot p(t)
\end{equation}
where $u_d$ is the unit color vector of the skin-tissue; $d_0$ is the stationary reflection strength; $u_p$ is the relative pulsatile strengths caused by hemoglobin and melanin absorption; $p(t)$ represents the physiological changes.
\begin{equation}
    v_s(t) = u_s \cdot (s_0 + \Phi(m(t),p(t)))
\end{equation}
where $u_s$ denotes the unit color vector of the light source spectrum; $s_0$ and $\Phi(m(t),p(t))$ denote the stationary and varying parts of specualr reflections; $m(t)$ denotes all the non-physiological variations such as flickering of the light source, head rotation, and facial express.

Therefore, performing diff operations between adjacent video frames can quantify the optical changes caused by physiological or non-physiological components.

\subsection{Network Structure}
We propose fast long-range spatio-temporal heart rate recognizing module in \ref{sec:3.4}. We provide some additional details on the specific implementation here.
\begin{itemize}
    \item Incorporating Temporal Shift to 2D Convolution for Lightweight Spatio-temporal Modeling. There are two method to use temporal shift module: In-place shift and Residual shift. We use Residual shift as it can address the degraded spatial feature learning problem, as all the information in the original activation is still accessible after temporal shift through identity mapping.
    \item Network structure of multi-branch parallel spatio-temporal model. The computational process for the djacent frame branch and the segment branch is the same. The input data size is 3x36x36. Performing the first convolution operation, the output feature map size is 32x36x36. Performing the second convolution operation, the output feature map size is still 32x36x36. Then perform average pooling and Dropout operations, with a pooling kernel size of 2x2 and a stride of 2, the output feature map size is 32x18x18. Next, perform one convolution operation, the output feature map size is 32x18x18. Then perform another convolution operation, the output feature map size is 64x18x18. Finally, perform one average pooling and Dropout operation, with a pooling kernel size of 2x2 and a stride of 2, the output feature map size is 64x9x9.
\end{itemize}

\subsection{Deployment} 
The CPUs of the 4 devices used in the experiments in \ref{sec:evaluation} are all ARM architecture, so we take Device $D_1$: Raspberry Pi 4B as an example to give the configuration required for deploying \sysname. We use the official Raspberry Pi system and install the driver for the test camera as the basic environment and use virtualenv as the Python environment management tool. The specific Python environment configuration is as follows: We run the program using Python 3.8.8, and use SciPy 1.8.0 and PyTorch 1.12 (CPU version) for network deployment and inference, and use OpenCV-Python 4.5.2 to process the video frames captured by the camera. The specific operations are as follows: First, please make sure that your device has completed the correct installation of the above environment and correctly set up the camera. Second, modifying the configuration file corresponding to \sysname so that the system can correctly read the input video stream and save the output results. Third, please use the recorded video file to test \sysname to ensure that the model can be read correctly and the system can run normally. Finally, please modify the system code to access the camera as real-time input. In addition, if necessary, you can configure the corresponding node.js and npm tools to draw the results in real time on a local web page.

When \sysname is performing inference, we use the dstat tool to monitor and record the real-time hardware usage of the device, such as CPU and memory. As for the real-time inference results, we use the charting tools in JavaScript to plot the real-time heart rate results inferred by UbiHR and present them on the frontend interface.

\color{black}